\documentclass[aps,prl,twocolumn,groupedaddress,showpacs]{revtex4-1}

\usepackage{graphicx,amsmath,multirow,hyperref,booktabs,units,amssymb,float}

\newcommand\solphys{\rmfamily{Sol.\ Phys.}}%
\newcommand\ssr{\rmfamily{Space Sci.\ Rev.}}%
\newcommand\grl{\rmfamily{Geophys. Res.\ Lett.}}%

\begin{document}


\title{Using the dipolar and quadrupolar moments to improve solar cycle predictions based on the polar magnetic fields}



\author{Andr\'es Mu\~noz-Jaramillo}
\email[]{amunoz@cfa.harvard.edu}
\affiliation{Harvard-Smithsonian Center for Astrophysics, Cambridge, MA 02138, USA}
\affiliation{University Corporation for Atmospheric Research, Boulder, CO 80307, USA}
\affiliation{Department of Physics \& Astronomy, University of Utah, Salt Lake City, UT 84112, USA}

\author{Laura A.\ Balmaceda}
\affiliation{Institute for Astronomical, Terrestrial and Space Sciences (ICATE-CONICET),San Juan, Argentina}

\author{Edward E. DeLuca}
\affiliation{Harvard-Smithsonian Center for Astrophysics, Cambridge, MA 02138, USA}

\date{\today}

\begin{abstract}
The solar cycle and its associated magnetic activity are the main drivers behind changes in the interplanetary environment and the Earth's upper atmosphere (commonly referred to as space weather and climate).  In recent years there has been an effort to develop accurate solar cycle predictions, leading to nearly a hundred widely spread predictions for the amplitude of solar cycle 24.  Here we show that cycle predictions can be made more accurate if performed separately for each hemisphere, taking advantage of information about both the dipolar and quadrupolar moments of the solar magnetic field during minimum.

\end{abstract}


\maketitle


The solar magnetic cycle is a process that brings the global magnetic field of the Sun (back and forth) from a configuration that is predominantly poloidal (contained inside the meridional plane), to one predominantly toroidal (wrapped around the axis of rotation; locally perpendicular to the meridional plane). During the first part of this process (poloidal to toroidal field), the poloidal components of the magnetic field are stretched and amplified by solar differential rotation\cite{parker1955a}.  This forms belts of amplified toroidal field which are transported to low latitudes, become buoyantly unstable due to overshooting convection, and rise to the surface to form bipolar sunspot groups (BSGs)\cite{parker1955b,fan2009}.  There are several mechanisms which may be playing a role during the second part of the process (toroidal to poloidal field)\cite{charbonneau2010} and the main contending theory at present is commonly referred to as the Babcock-Leighton (BL) mechanism\cite{babcock1961,leighton1969}:   The fact that BSGs present a systematic tilt with respect to a line parallel to the solar equator\cite{hale-etal1919} in combination with surface processes of diffusion and advection, has as a consequence a net transport of flux towards the poles that cancels the old polarity and reverses the sign of the poloidal field, setting the stage for the following cycle \cite{babcock1961,leighton1969,wang-nash-sheeley1989b}.

Due to its cyclic modulation of the heliospheric environment \cite{schwenn2006}, the Earth's magnetosphere \cite{pulkkinen2007}, and the Sun's radiative output \cite{domingo-etal2009}, the prediction of the solar cycle has commanded an increasingly large effort since the dawn of the space age \cite{petrovay2010}.  Cycle predictions are typically classified into extrapolation methods, which use the mathematical properties of the sunspot data series to predict future levels of activity; precursor methods, which use different measurable quantities as a proxy to estimate the subsequent cycle's amplitude; and model-based predictions which use the assimilation of data into models of the solar cycle to make predictions.  There is, however, no consensus yet about the most effective method of cycle prediction, evidenced by nearly a hundred widely spread predictions for the amplitude of solar cycle 24 (whose prediction range spans all cycle amplitudes ever observed) \cite{petrovay2010,pesnell2012}.

\begin{figure*}
\includegraphics[width=\textwidth]{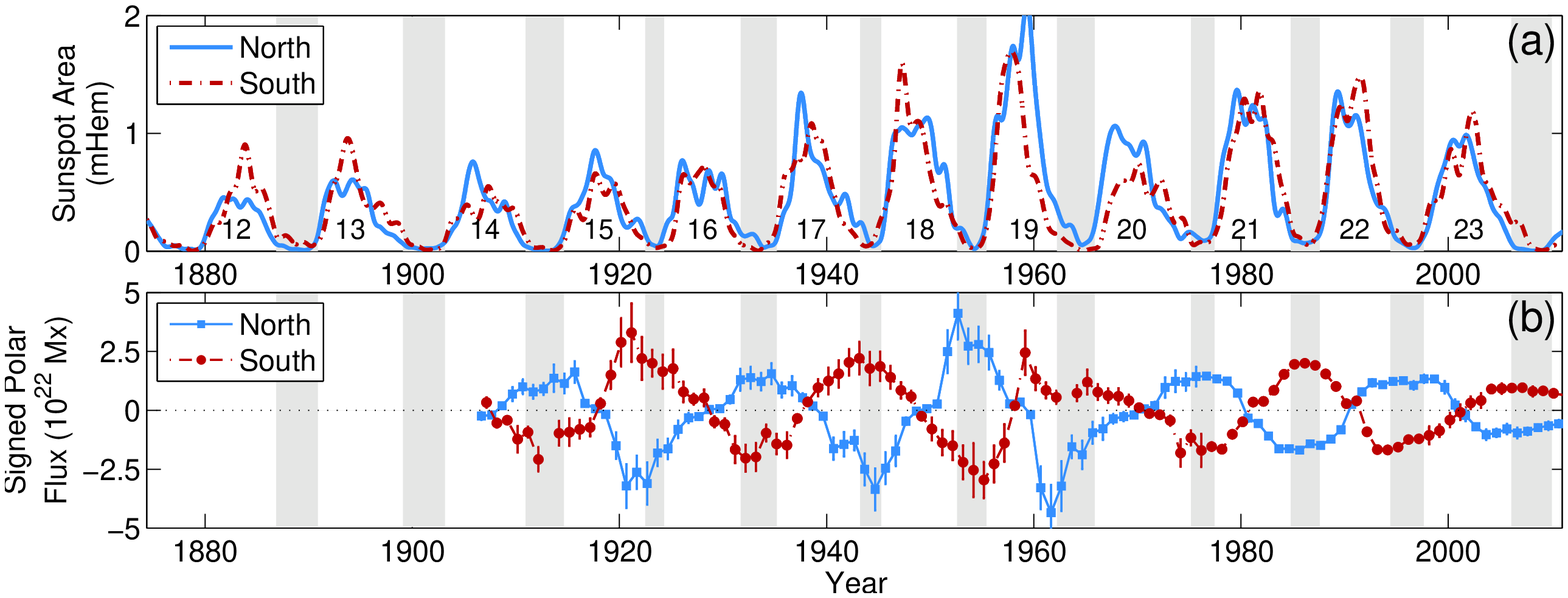}
\caption{(a) Smoothed daily sunspot area for the northern (solid blue line) and southern (dashed red line) hemispheres. (b) Polar flux (based on magnetic and polar faculae observations) for the northern (blue squares) and southern (red circles) hemispheres.  Shaded areas indicate the duration of solar minimum defined as the period between points set at 15\% of the amplitude of the corresponding bracketing cycle.  Unless otherwise noted, all polar flux values used in this letter correspond to minimum averages. \label{Fig_data}}
\end{figure*}

One of the determinant factors shaping the nature of current prediction methods is the availability (or lack) of long-term solar records.   For example, while most precursor methods are based on the logic that polar fields at solar minimum are the seed of the following cycle (first used to predict solar cycle 21 \cite{schatten-etal1978}), in reality most use geomagnetic activity measurements for predictions \cite{pesnell2012} due to the lack of polar field measurements before 1970.  Another important limiting factor arises from the fact that both the sunspot record -- which has long been regarded as one of the main indicators of solar activity and thus is used by most to calibrate and verify cycle prediction -- and geomagnetic activity are solar global variables.  This has resulted in cycle predictions dealing exclusively with the whole-Sun cycle amplitude while, in reality, hemispheric asymmetries of both the sunspot record \cite{zolotova-etal2010,norton-gallagher2010} and the polar fields \cite{svalgaard-kamide2012} suggest that the cycle in the northern and southern hemispheres are loosely coupled and should be predicted separately.

In this letter we take advantage of a recently standardized database of polar faculae measurements going back to the beginning of the 20th century (as a proxy for the evolution of the polar magnetic flux)\cite{munoz-etal2012b}, in combination with a long-term homogeneous sunspot area database\cite{balmaceda-etal2009}, to demonstrate the advantages of using the dipolar and quadrupolar moments of the solar magnetic field to make hemispheric predictions.  Additionally, by extending the observed relationship between the polar field and the amplitude of the next cycle to a full century, we substantiate predictions based on the polar field \cite{schatten2005, svalgaard-etal2005,choudhuri-chatterjee-jiang2007} -- currently inconspicuous among the many different predictions of solar cycle 24.

\section{Data}

In this work we use a homogeneous database of sunspot areas \cite{balmaceda-etal2009}, separated in northern and southern hemisphere sets, calculating the total hemispheric daily sunspot area (Fig.~\ref{Fig_data}-a).  Area belonging to groups observed at the equator are not assigned to any of the two hemispheres.  We remove high-frequency components by convolving our data series with a modified 24-month Gaussian filter \cite{hathaway2010}, found to yield more consistent results while finding maxima and minima using different activity proxies than the traditional 13-month running mean.

Our magnetically calibrated polar faculae database (Fig.~\ref{Fig_data}-b) comes from a recent calibration and standardization \cite{munoz-etal2012b} of four facular Mount Wilson Observatory (MWO) data reduction campaigns \cite{sheeley1966,sheeley1976,sheeley1991,sheeley2008}.   Consecutive campaigns were cross-calibrated using five year overlaps and validated using an automatic detection algorithm on intensity data from the Michelson Doppler Imager\cite{scherrer-etal1995}.  The resultant faculae database was calibrated in terms of polar magnetic field and flux using magnetic field measurements taken by the Wilcox Solar Observatory and SOHO/MDI (see Supplemental Material for more details on our datasets).

\begin{figure*}
\includegraphics[width=\textwidth]{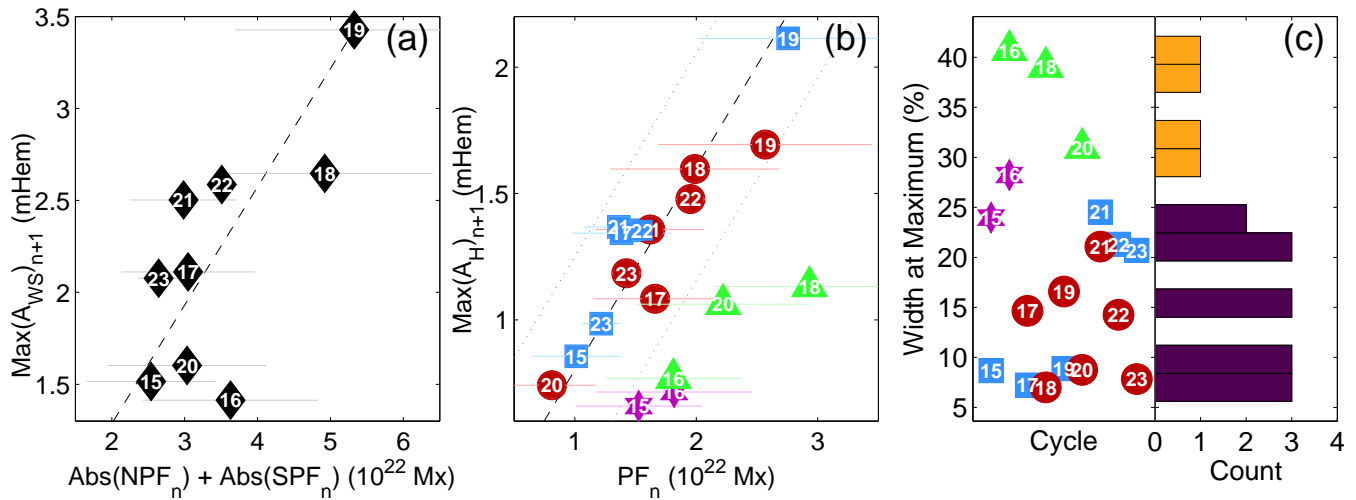}
\caption{(a) Average of the unsigned northern (NPF) and southern (PFN) polar fluxes (as an indicator of the Sun's dipole moment) at the minimum of cycle $n$ vs.\ amplitude of the next whole-Sun cycle. Numbers denote the cycle being predicted. (b) Polar fluxes (PF) for the northern (blue squares and green triangles) and southern (red circles and magenta stars) hemispheres at the minimum of cycle $n$  vs.\ amplitude of the next corresponding hemispheric cycle.  Error bars are shown as faint horizontal lines.  The dashed line in both panels corresponds to a linear fit using the least absolute residuals method. Numbers indicate the cycle being predicted. (c) Histogram of hemispheric width at maximum in units of cycle length.  Peaked hemispheric cycles are denoted using blue squares (red circles) for the northern (southern) hemisphere.  Cycles with an extended maximum are denoted using green triangles (magenta stars) for the northern (southern) hemisphere; markers in all hemispheric scatter-plots have this same meaning. Histogram bars are colored to separate the bins which contain cycles belonging to the two different branches.\label{Fig_Cycles}}
\end{figure*}

\section{Hemispheric vs.\ Whole-Sun Cycles}

Following the current standard practice of making whole-Sun predictions, our first task is to study the relationship between the Sun's axial dipole moment at minimum (which is proportional to the unsigned average of the northern and southern polar magnetic fluxes) and the amplitude of the next cycle.  We find them to be correlated (with a with a Pearson's correlation coefficient of $\rho = 0.69$ and $P=96\%$ confidence level; see Fig.~\ref{Fig_Cycles}-a).  However, a linear fit using least absolute residuals (LAR; which naturally gives less weight to possible outliers in the dataset), shows a departure from the linear relationship one expects from the amplification of toroidal field out of poloidal field by differential rotation (an issue that does not affect cycles so far predicted using polar field measurements, i.e.\ 21-23).  This deviation from linearity becomes more evident while looking at it from a hemispheric point of view (Fig.~\ref{Fig_Cycles}-b), where a linear fit using LAR highlights the apparent existence of two separate branches.  A comparison between the hemispheric and whole-Sun relationship shows that deviations using whole-Sun cycles are associated with a hemisphere falling outside the main branch.

\subsection{Solar Magnetic Moments and their Relationship with Irregularities in Cycle Shape}

A qualitative assessment of hemispheric cycles and polar fluxes during the preceding minimum (Figs.~\ref{Fig_data}-a \& b) shows that off-branch hemispheric cycles (15S, 16N, 18N and 20N, shown in Fig.~\ref{Fig_Cycles}-b with triangular and star markers) are characterized by an extended multimodal maximum (as opposed to hemispheric cycles in the main branch, which generally show a peaked shape)-- a characteristic that we quantify by dividing the cycle into rising, maximum and decay phases and measuring the duration (width) of the maximum phase (see Supplemental Material).  Additionally, we find off-branch hemispheric cycles to be preceded by minima characterized by magnetic flux imbalance between the north and the south poles.  Note that these cycles correspond to cycles for which only facular data is available, so we cannot rule out completely that these imbalances are caused by issues in the facular data.  However, the strongest polar flux asymmetry in our dataset (taking place around 1960) is also visible using MWO magnetograms as well (Leif Svalgaard, private communication).  This suggests that these asymmetries are real.

\begin{figure*}
\includegraphics[width=\textwidth]{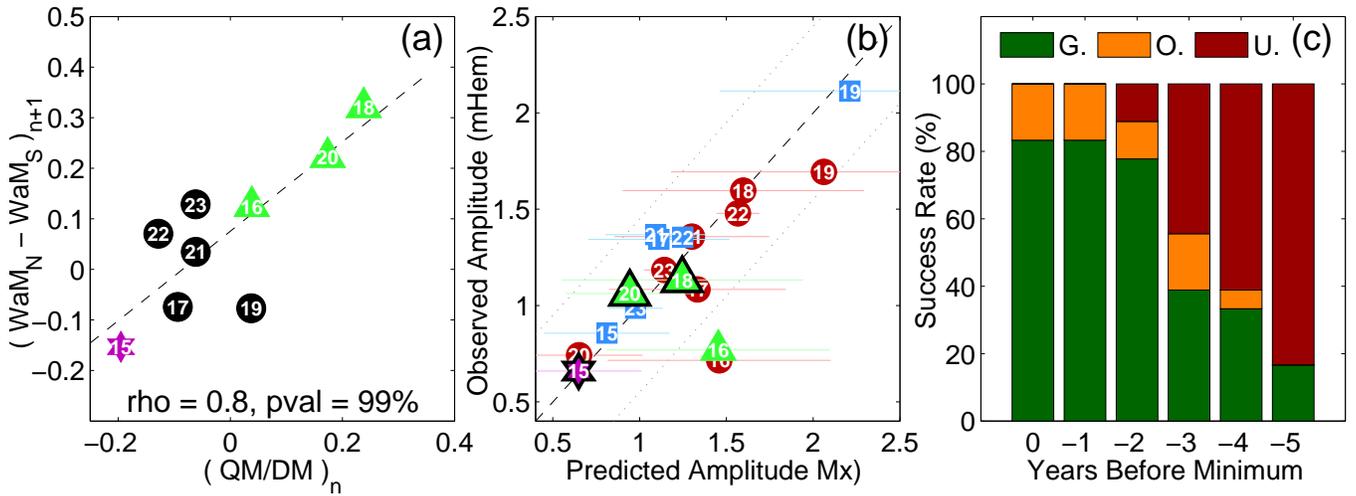}
\caption{(a) Ratio between the solar dipolar (DM) and quadrupolar (QM) moments at the minimum of cycle $n$ vs.\ difference in cycle width at maximum. (b) Predicted vs observed cycle amplitude. Cycles with $|QM/DM|\geq 16.5\%$ are predicted using the off-branch relationship (denoted with black outlines). Error bars are shown as faint horizontal lines. (c) Success rate of the prediction method when made using polar flux measurements taken at, or 1-5 years before minimum.  The lower section of each column (dark green) indicates predictions within the 99\% confidence bounds, the middle section (light yellow) overestimated amplitudes, and the top section (red) underestimated amplitudes. See Supplemental Material for the scatter plots used to create this figure.\label{Fig_Irr}}
\end{figure*}

A histogram of hemispheric cycle \emph{width at maximum} (WaM) (Fig.~\ref{Fig_Cycles}-c) shows how off-branch hemispheric cycles are consistently those with the highest values. Considering that sunspot cycles in our dataset generally have only one off-branch hemisphere (or none), this means that sunspot cycles with hemispheres in separate branches are characterized by a strong asymmetry in shape.  This hemispheric asymmetry is well correlated with the relative strength of the axial quadrupolar (QM) and dipolar (DM) moments during the preceding minimum (with a Pearson's correlation coefficient of $\rho = 0.8$ and $P=99\%$ confidence level; Fig.~\ref{Fig_Irr}-a), calculated using the difference and average, respectively, of the unsigned northern and southern polar fluxes (see Supplemental Material for more details).

Invoking recent high-resolution observations of the polar field, showing it to be concentrated in unipolar patches of magnetic field (in many cases of mixed opposite polarities) \cite{shiota-etal2012}, we can propose a possible explanation of the relationship between a significant QM and hemispheric asymmetry: a significant QM means that while in one polar crown almost all poloidal field bundles are of the same polarity, in the other one there is a higher mixture of patches with opposite polarities.  These conflicting bundles are wound independently by differential rotation, cancelling and interacting with each other as the cycle progresses, resulting in a multimodal hemispheric cycle (with a lower amplitude than a smooth cycle would have), while in the other hemisphere the cycle turns out nice and sharp.

\section{Prediction of Hemispheric cycles}

We refine predictions based on the polar fields by taking advantage of the fact that $QM/DM$ during minimum is a good indicator of whether one (and which) of the subsequent hemispheric cycles will have an extended maximum (and thus be off the main branch).  We perform separate fits to the main and secondary branches (shown in Fig.~\ref{Fig_Cycles}-b) and use an upper (lower) limit of $QM/DM\geq lim=16.5\%$ ($QM/DM\leq lim=-16.5\%$) as criteria for choosing the relationship used for prediction of the northern (southern) hemispheric cycle.  Our predictors become:

\begin{equation}\label{Eq_Preda1}
   \operatorname{Amp}(\operatorname{PFN})_{n+1} =  \left\{\begin{array}{cc} a_{mb}\operatorname{PFN}_n & \frac{QM}{DM} \leq lim\\\\
                                                                            a_{sb}\operatorname{PFN}_n & \frac{QM}{DM} > lim
                                                    \end{array}\right.
\end{equation}
and
\begin{equation}\label{Eq_Preda2}
   \operatorname{Amp}(\operatorname{PFS})_{n+1} =  \left\{\begin{array}{cc} a_{mb}\operatorname{PFS}_n & \frac{QM}{DM} \geq -lim\\\\
                                                                            a_{sb}\operatorname{PFS}_n & \frac{QM}{DM} < -lim
                                                    \end{array}\right.,
\end{equation}
where $a_{mb}=0.802$ mHem/10$^{22}$Mx ($a_{sb}=0.425$ mHem/10$^{22}$Mx) is the proportionality coefficient of the main (secondary) branch.

Considering that there is not a significant quadrupolar moment during the minimum of sunspot cycle 23 ($QM/DM=0.05$), we use the main branch's relationship to predict an amplitude of $590\pm143$ $\mu$Hem (sunspot number $R=36\pm9$) for the northern hemisphere and $664\pm108$ $\mu$Hem (sunspot number $R=41\pm7$) for southern hemisphere in cycle 24 (Fig.~\ref{Fig_Irr}-b).  Together they give a maximum of $1254\pm251$ $\mu$Hem (sunspot number $R=77\pm16$) for the amplitude of cycle 24, making cycle 24 one of the weakest cycles in the last hundred years, agreeing with other predictions based on the solar polar field \cite{schatten2005,svalgaard-etal2005}.

To finalize, we study the efficacy of hemispheric predictions using polar flux measurements taken at, and before solar minimum. Fig.~\ref{Fig_Irr}-c shows a quantitative assessment of this performance in time.  We consider the prediction to be accurate if it differs from the observed amplitude by less than our fit's $99\%$ confidence bounds.  In particular, we find predictions for solar cycle 24 to change only by $10\%$ during the three years before minimum (from sunspot number $R=85\pm10$ using values from 2005 to $R=77\pm16$ at solar minimum in 2008); however, most minima in our database do not seem to stabilize as early. We find the method to perform well up to two years before minimum (with a success rate of $83-78\%$), after which the success rate drops dramatically.

\section{Concluding Remarks}

The results presented here (involving a full century of observations) demonstrate the power of solar polar fields during solar minimum as predictors of the amplitude of the next cycle  (and do so in agreement with our theoretical understanding of the solar cycle).  In particular we show how polar flux becomes a better cycle predictor by taking advantage of the hemispheric polar fields to calculate both the dipolar and quadrupolar moments -- the reason being that minima with significant quadrupolar moments lead to irregular hemispheric cycles with lower effective amplitudes than they would have if they were not irregular.  We predict smooth hemispheric cycles for solar cycle 24 with amplitudes of $R=36\pm9$ ($R=41\pm7$) for the northern (and southern) hemispheres for a total whole-Sun amplitude of $R=77\pm16$.

Our work paves the way for a new generation of precursor methods where the objective is no longer to find which variable yields the most accurate predictions, but rather how to make predictions better.  One of the crucial points that needs to receive more attention is the timing of the solar cycle, both in term of solar maximum (which is as important for long term planning as cycle amplitude) and solar minimum (considering that predictions based on the polar field are only accurate if made within two years of minimum).  Another important issue is to broaden the concept of cycle prediction to include solar minimum conditions; in order to extend our predictive capability in time (ideally to more than one solar cycle).

Above all, our results add to the mounting evidence showing the solar poles to be a crucial link in the evolution of the solar cycle.  We anticipate that Solar Orbiter, an ESA mission under development, by going out of the ecliptic and looking down on the poles will be able to uncover unknown details of the polar magnetic field evolution, thus considerably enhancing its understanding in the coming decade -- specially in conjunction with the long-term full-Sun view of NASA's Solar Dynamics Observatory and the high-resolution observations of Solar-C of ISAS/JAXA.

\begin{acknowledgments}
We are thankful to Sami Solanki, Mar\'ia Dasi-Espuig and Mar\'ia Navas-Moreno, and two anonymous referees for useful discussions and suggestions. This research is supported by the NASA Living With a Star Jack Eddy Postdoctoral Fellowship Program, administered by the UCAR Visiting Scientist Programs.  Andr\'es Mu\~noz-Jaramillo is grateful to David Kieda for his support and sponsorship at the University of Utah. E.\ DeLuca was supported by contract SP02H1701R from Lockheed Martin to SAO.
\end{acknowledgments}

\bibliography{References}

\clearpage
\onecolumngrid

\begin{figure}[H]
\centering
\begin{tabular}{c}
  \includegraphics[width=0.7\textwidth]{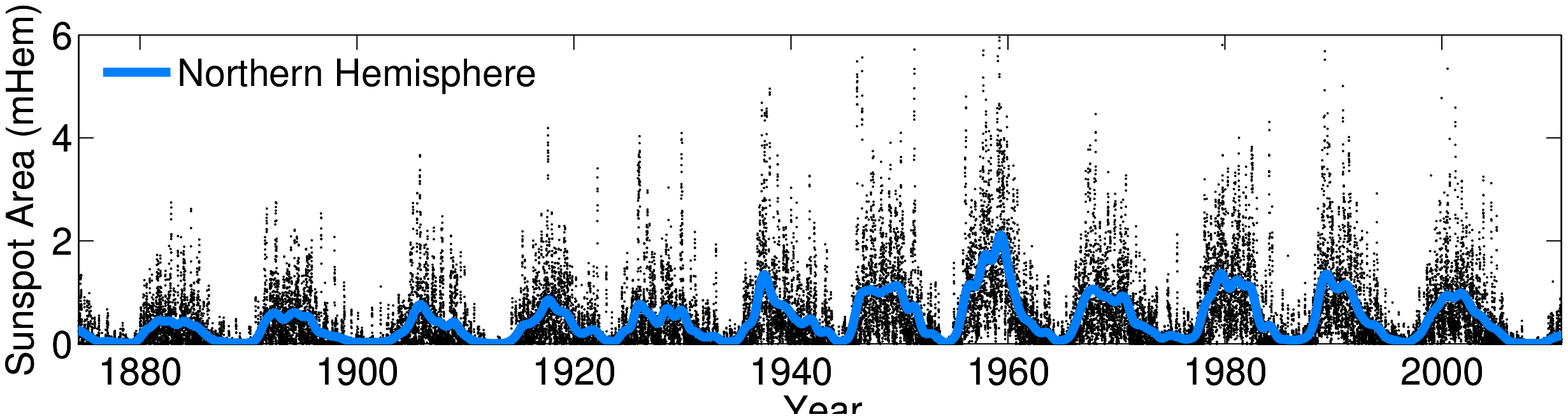}\\
  \includegraphics[width=0.7\textwidth]{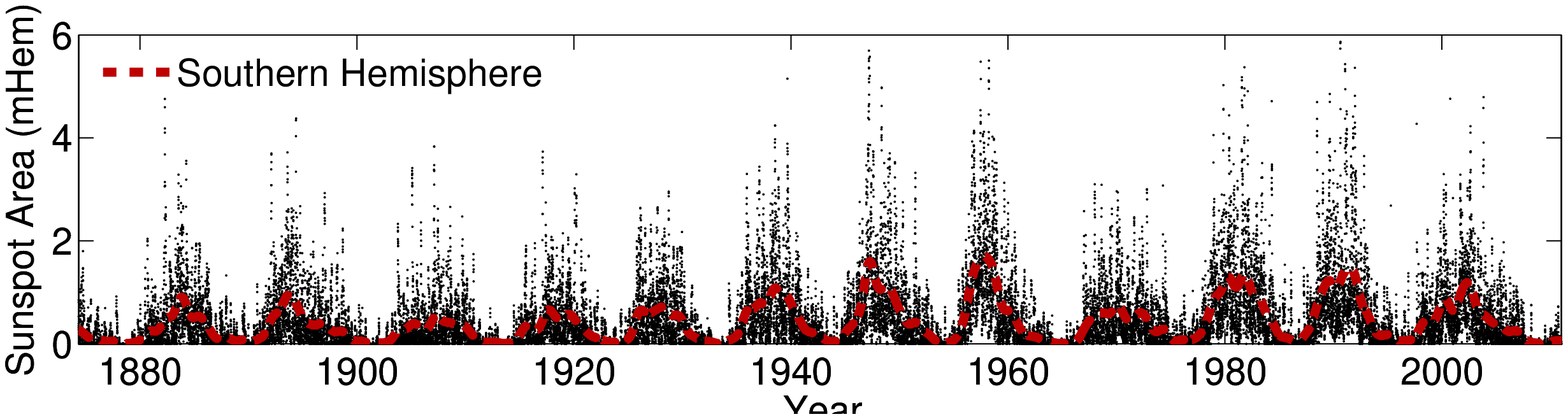}
\end{tabular}
\caption{Total daily sunspot area (black dots) is calculated for the northern (top panel) and southern (bottom panel) hemispheres.  A 24-month Gaussian filter is applied to remove the high-frequency component in the data series shown as a solid blue line for the northern (top panel) hemisphere and as a dashed red line for the southern (bottom panel) hemisphere.}\label{Fig_Dat}
\end{figure}

\section{SUPPLEMENTAL MATERIAL}

\section{Sunspot Area: Data, Smoothing, and Cycle Separation}

In this work we use a homogeneous database of sunspot areas mainly based on observations taken by the Royal Greenwich Observatory, several stations belonging to the former USSR (compiled in the \emph{Solnechniye Danniye} bulletin issued by the Pulkovo Astronomical Observatory), and the US Air Force Solar Optical Observing Network (SOON)\cite{balmaceda-etal2009}.  We separate the data in northern (top panel in Figure \ref{Fig_Dat}) and southern (bottom panel in Figure \ref{Fig_Dat}) hemisphere sets, calculating the total hemispheric daily sunspot area.  Area belonging to groups observed at the equator are not assigned to any of the two hemispheres.

\begin{figure}[!h]
\centering
\begin{tabular}{c}
  \includegraphics[width=.75\textwidth]{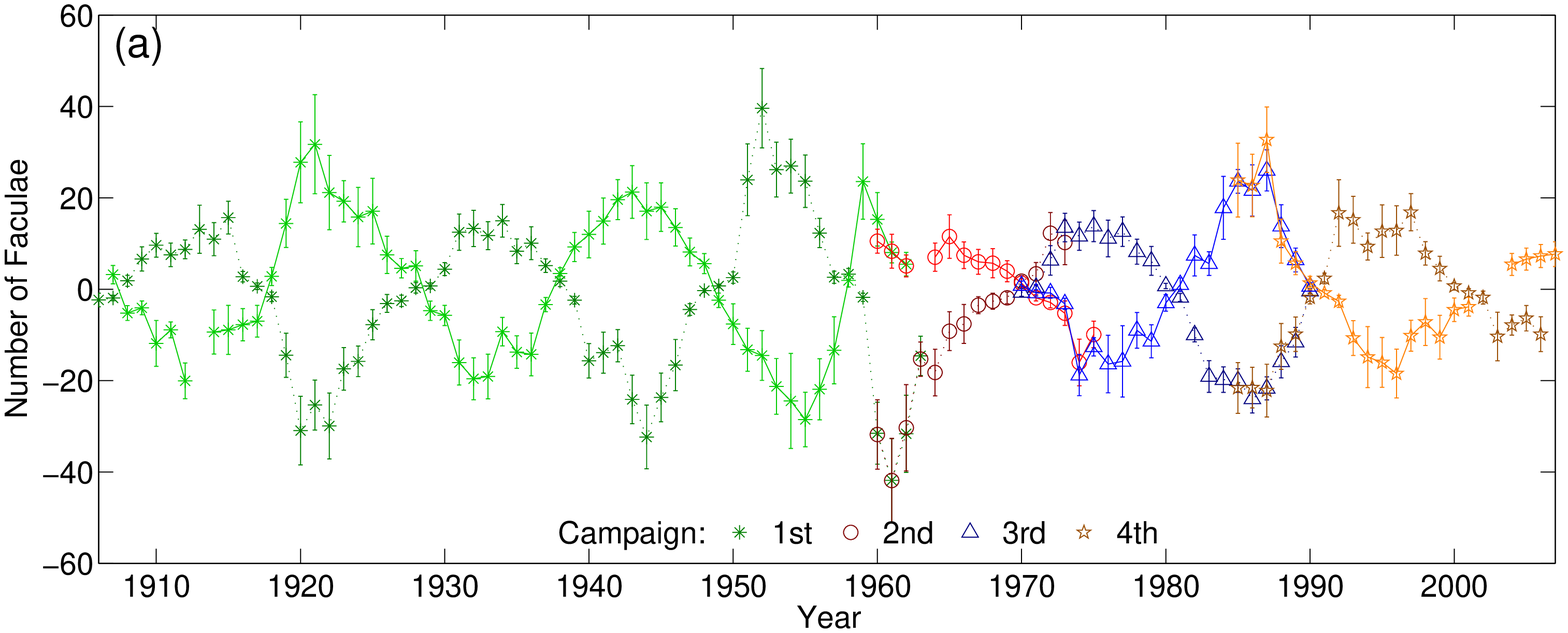}\\
  \includegraphics[width=.75\textwidth]{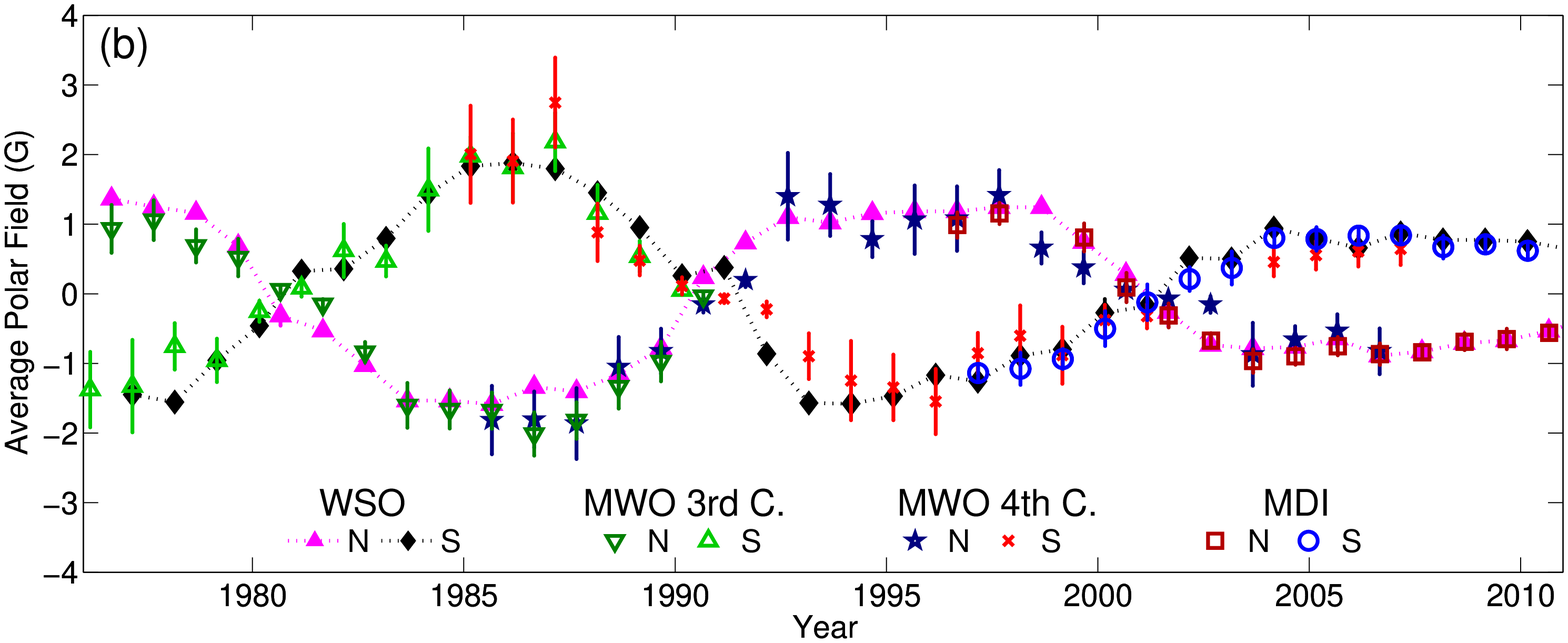}\\
  \includegraphics[width=0.8\textwidth]{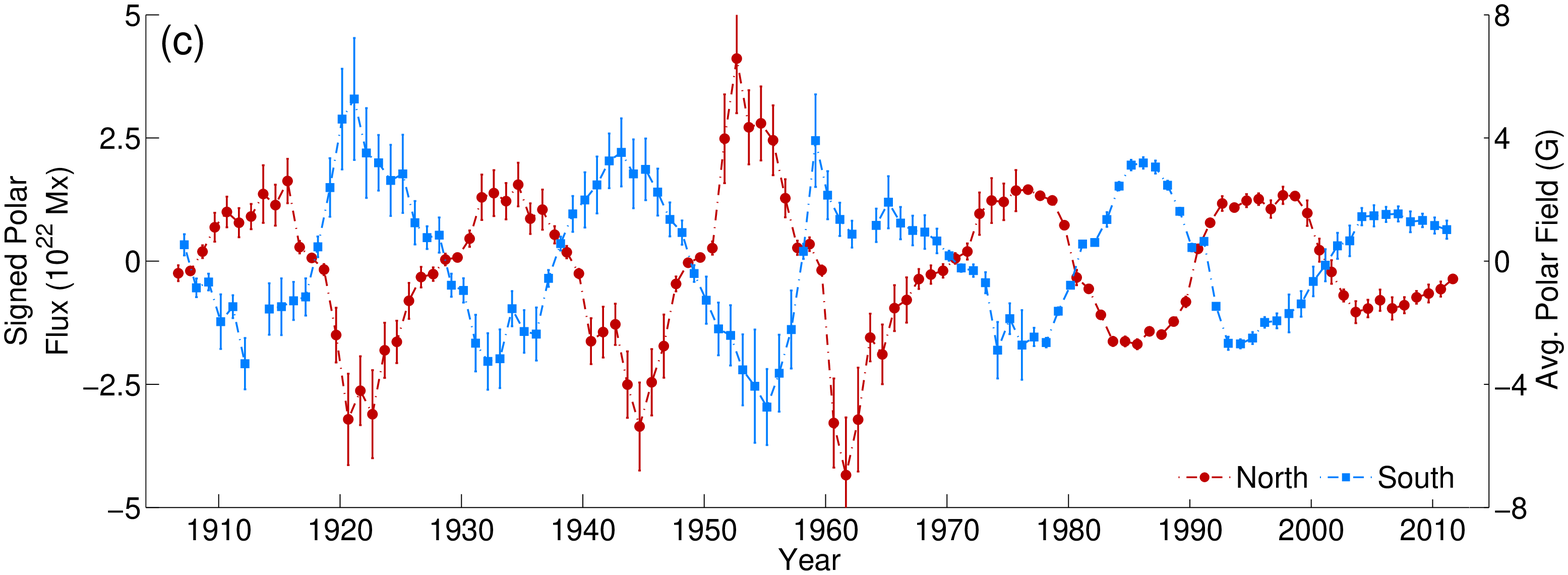}
\end{tabular}
\caption{Four Mount Wilson Observatory campaigns are standardized using their overlap to obtain a consistent polar faculae database (a).  Different colors and markers correspond to different data reduction campaigns.  This database is calibrated using data from the Wilcox Solar Observatory and the Michelson Doppler Imager on the SOlar and Heliospheric Observatory, in order into convert it to polar flux estimates (b).  The resultant databases are consolidated into a single proxy used in this work (c).}\label{Fig_PF_Dat}
\end{figure}

We remove high-frequency components by convolving our data series with the modified 24-month Gaussian filter \cite{hathaway2010}:
\begin{equation}\label{Eq_Filter}
        F(t,t',a) =\left\{\begin{array}{ccc}
                     0 & & t \leq t' - 2a\\
                     K e^{-\frac{(t - t')^2}{2a^2}} - e^{-2} \left(3 - \frac{(t - t')^2}{2a^2}\right) & & t' - 2a < t \leq t' + 2a\\
                     0 & & t > t' + 2a
               \end{array}\right.,
\end{equation}
where $t'$ denotes the position of the center and $a=12$ months the half-width of the Gaussian filter, and $K$ is a normalization constant which ensures that the integral of the filter is equal to one.  This type of filters has been found to yield more consistent results while finding maxima and minima (using different databases like the international sunspot number and the 10.7cm radio flux), than the traditional 13-month running mean.  After applying the filter we find the minima separating the different cycles from each other.  This is done independently for each hemisphere, which means that the time of minimum may be different in each hemisphere (in average six months apart).

\section{Polar Flux Data}

Our magnetically calibrated polar faculae database comes from the recent standardization \cite{munoz-etal2012b} of four Mount Wilson Observatory data reduction campaigns\cite{sheeley1966,sheeley1976,sheeley1991,sheeley2008}.   We took advantage of the five year overlap between consecutive campaigns to cross-calibrate different data reduction campaigns (see Figure \ref{Fig_PF_Dat}-a), and validated using an automatic detection algorithm on intensity data from the Michelson Doppler Imager\cite{scherrer-etal1995} (MDI).  The resultant database is then calibrated in terms of polar magnetic field and flux using magnetic field measurements taken by the Wilcox Solar Observatory and SOHO/MDI (see Figure \ref{Fig_PF_Dat}-c).   Once we convert polar facular count into polar flux values we combine MWO, WSO and MDI data into a single consolidated database (see Figure \ref{Fig_PF_Dat}-c).  This database has been already used to study the role of the polar magnetic flux in the evolution of the heliospheric magnetic field (HMF) at Earth by comparing it with HMF reconstructions spanning more than a century\cite{munoz-etal2012b}.  Note that even though we generally use the term ``polar flux" in the letter and supplemental material, data points for cycles 15-20 are based exclusively on magnetically calibrated facular measurements.

\clearpage

\section{Width at Maximum as Cycle Discriminator}

\begin{figure}
\centering
\begin{tabular}{ccc}
  \includegraphics[width=0.32\textwidth]{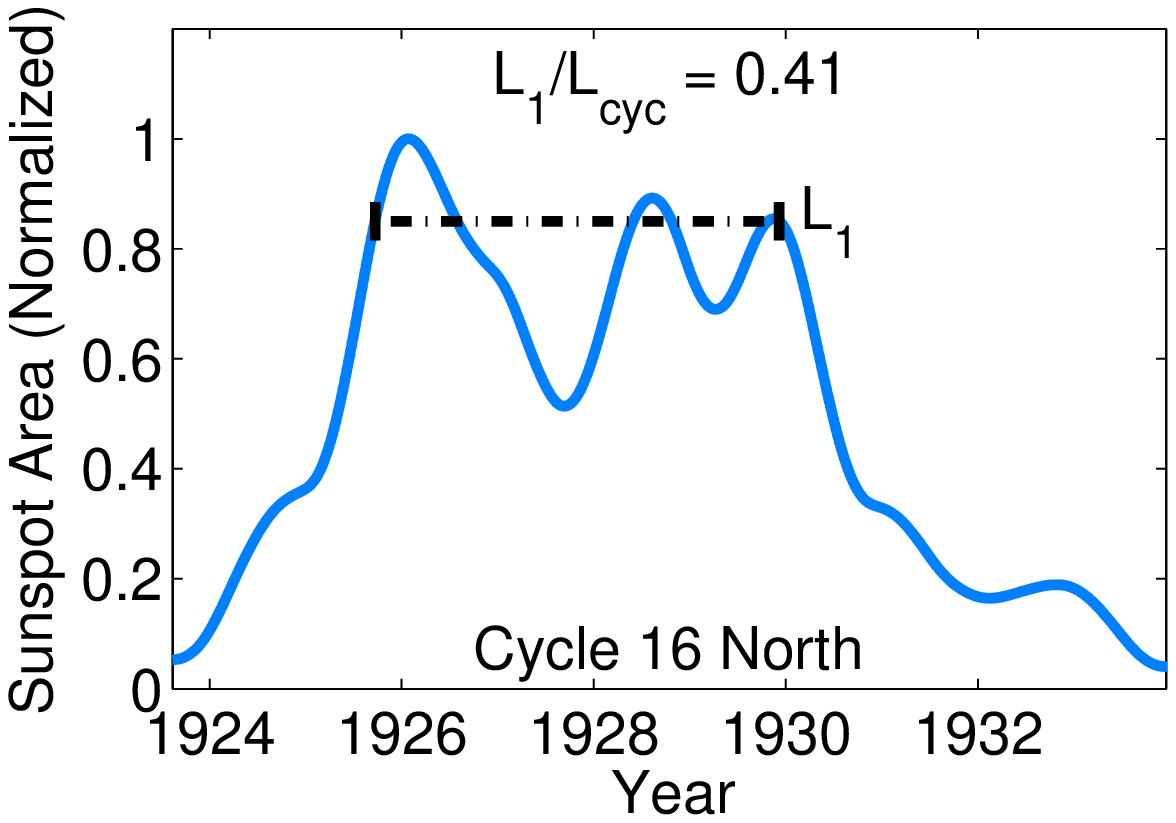} & \includegraphics[width=0.32\textwidth]{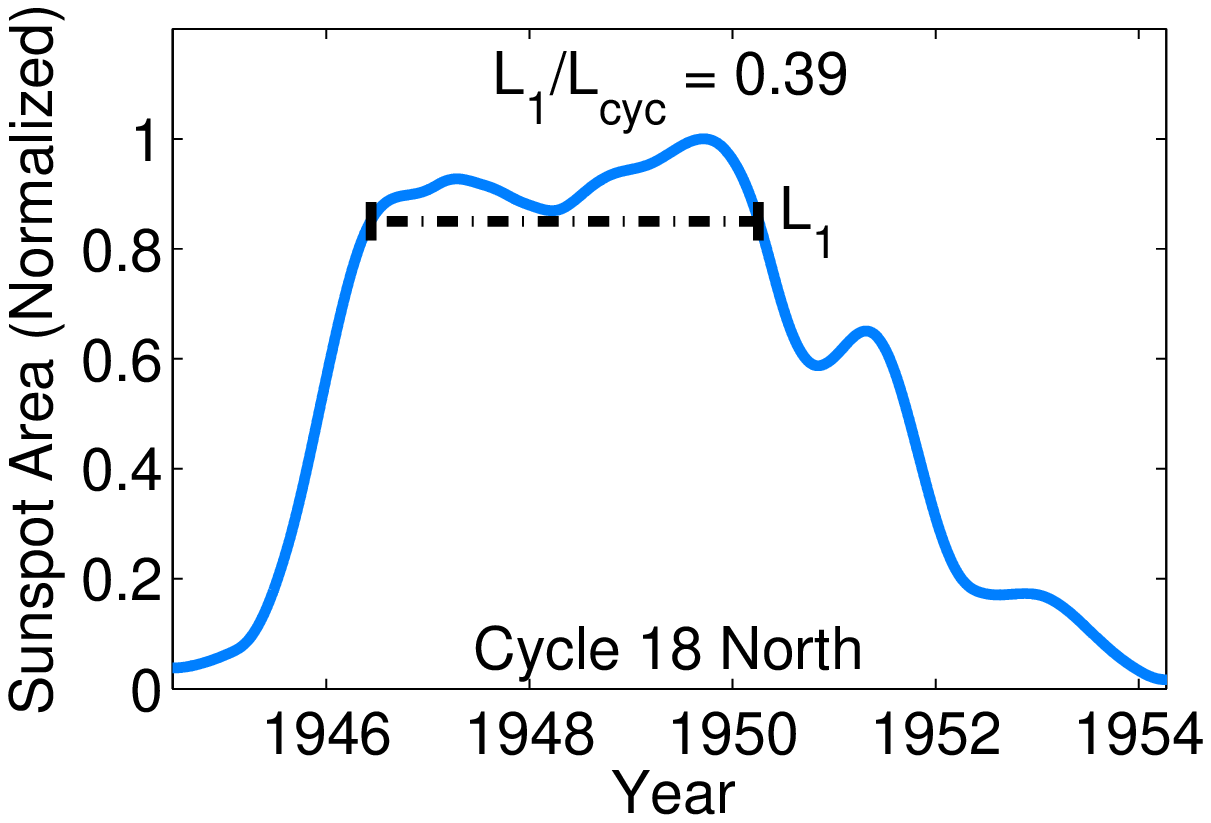} & \includegraphics[width=0.32\textwidth]{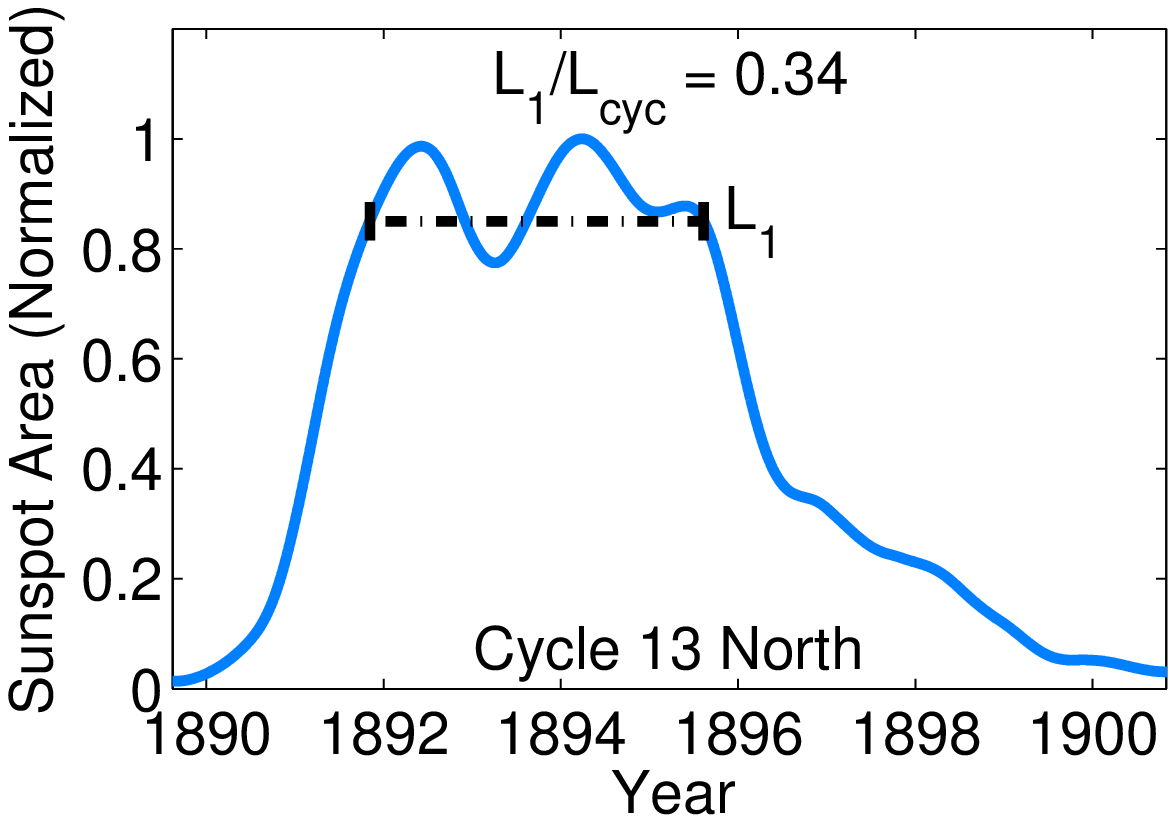}\\
  \includegraphics[width=0.32\textwidth]{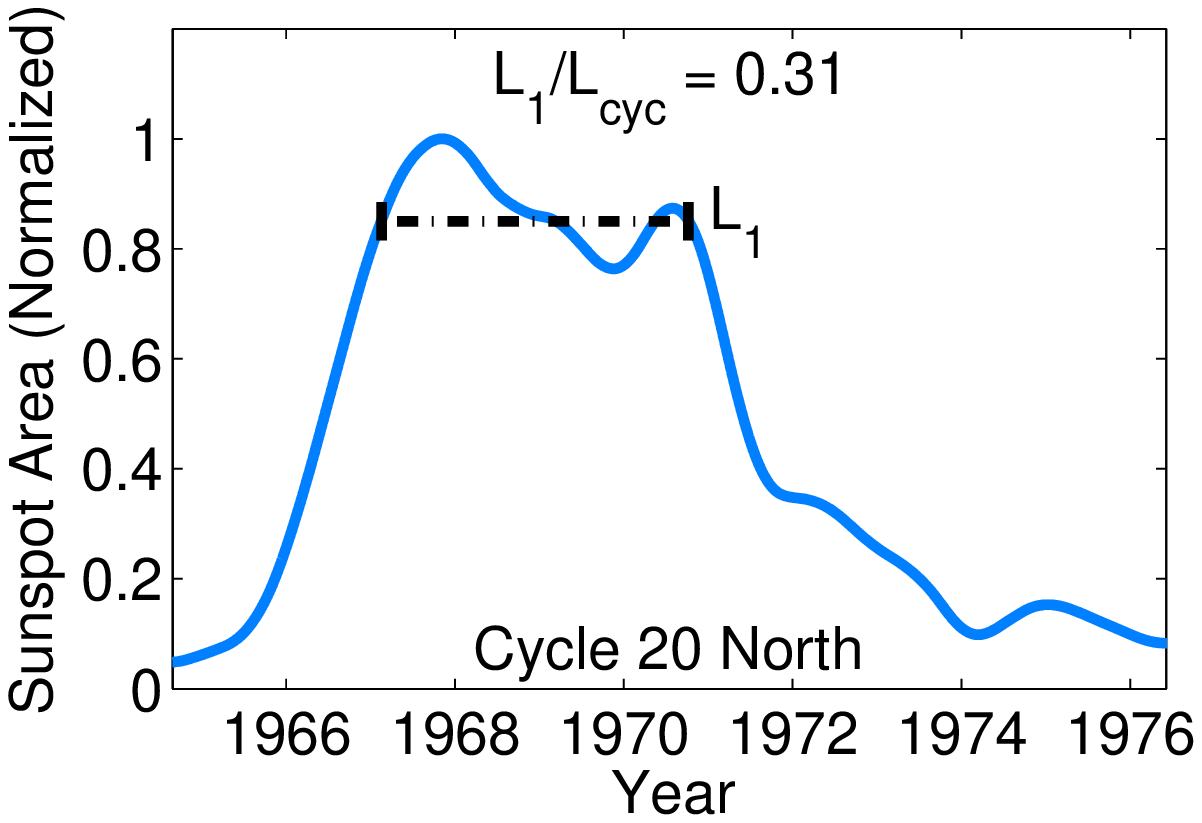} & \includegraphics[width=0.32\textwidth]{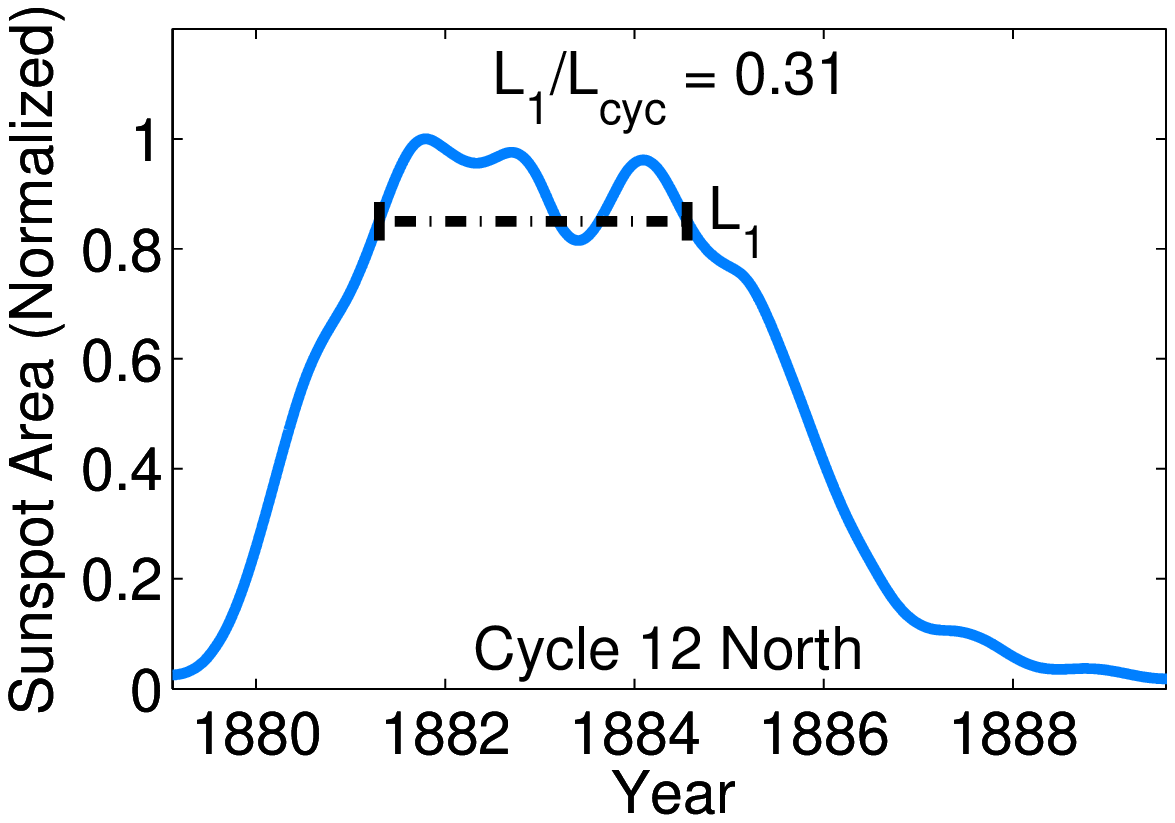} & \includegraphics[width=0.32\textwidth]{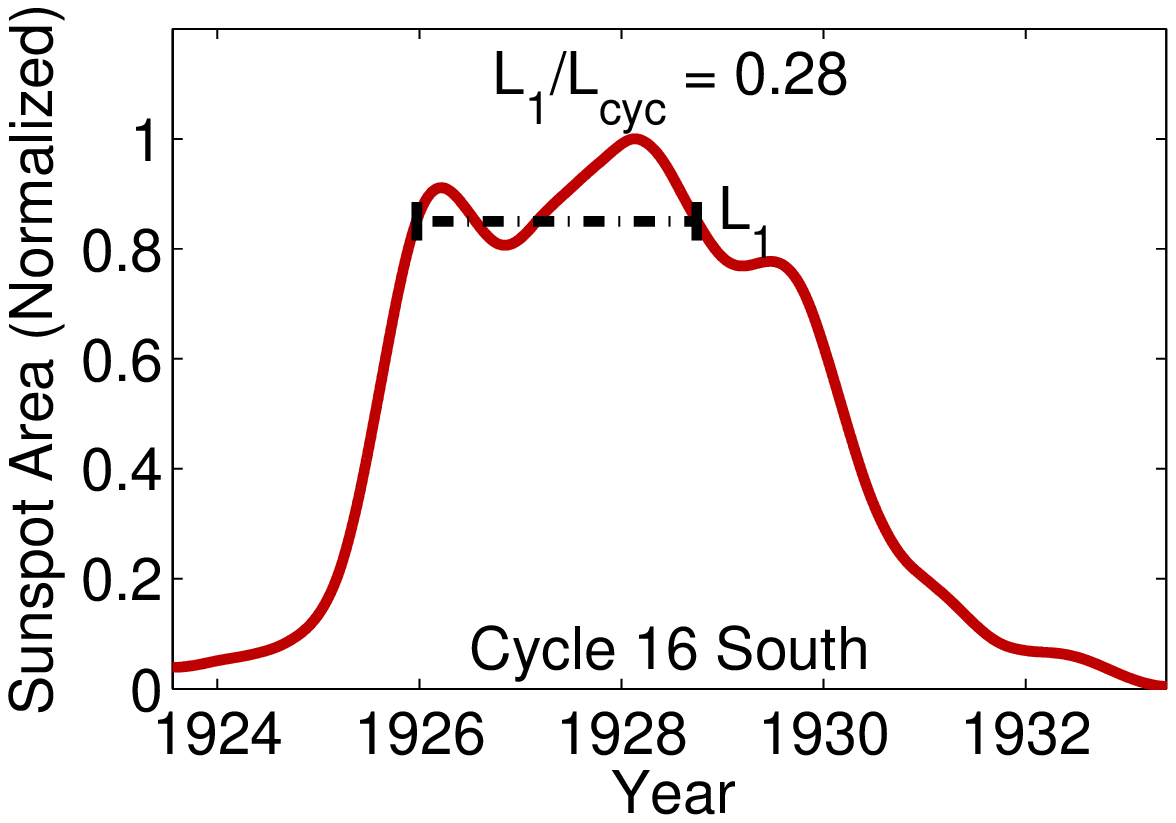}\\
  \includegraphics[width=0.32\textwidth]{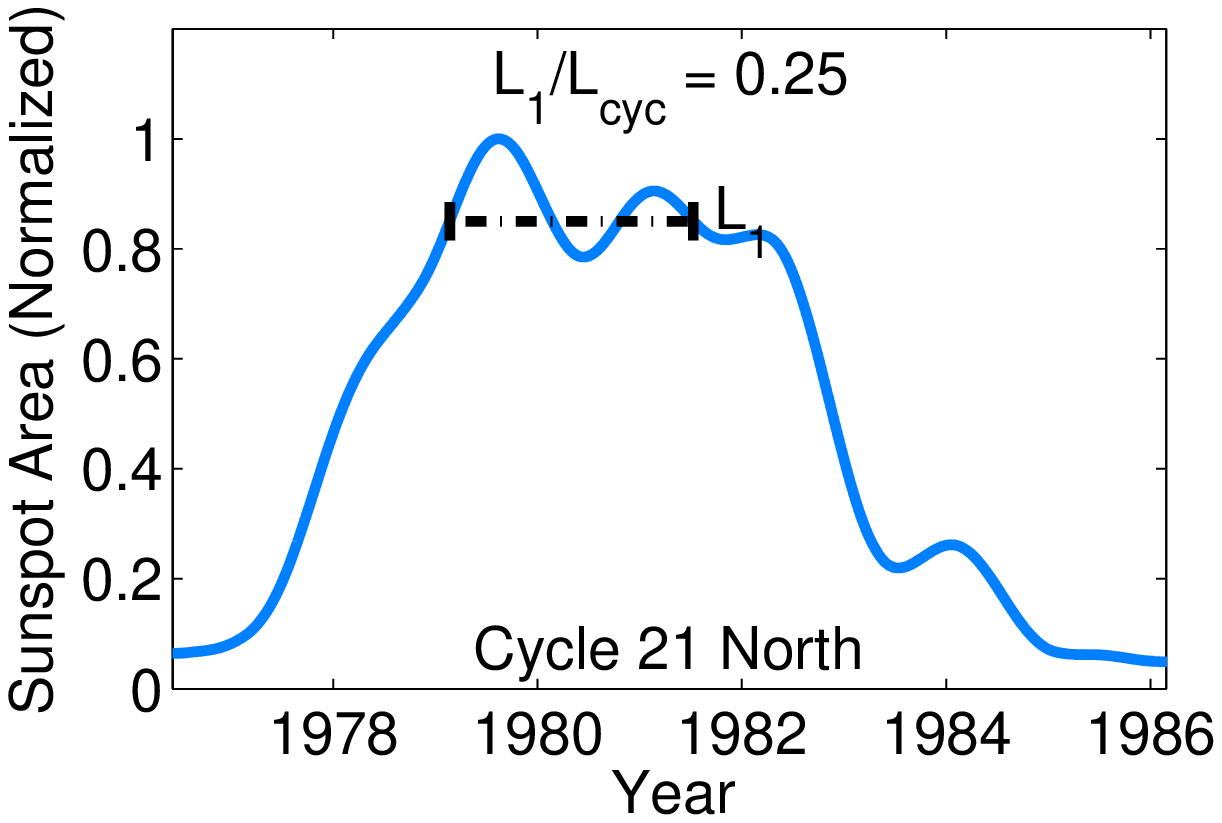} & \includegraphics[width=0.32\textwidth]{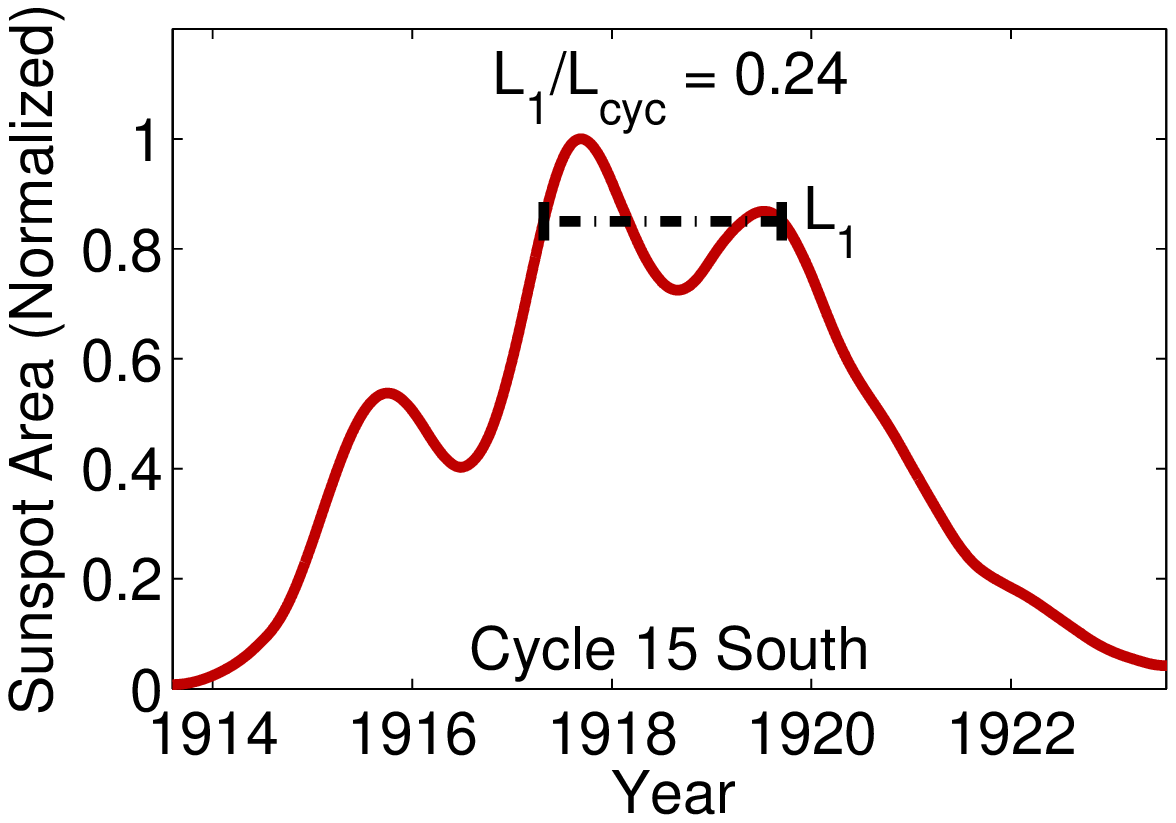} & \includegraphics[width=0.32\textwidth]{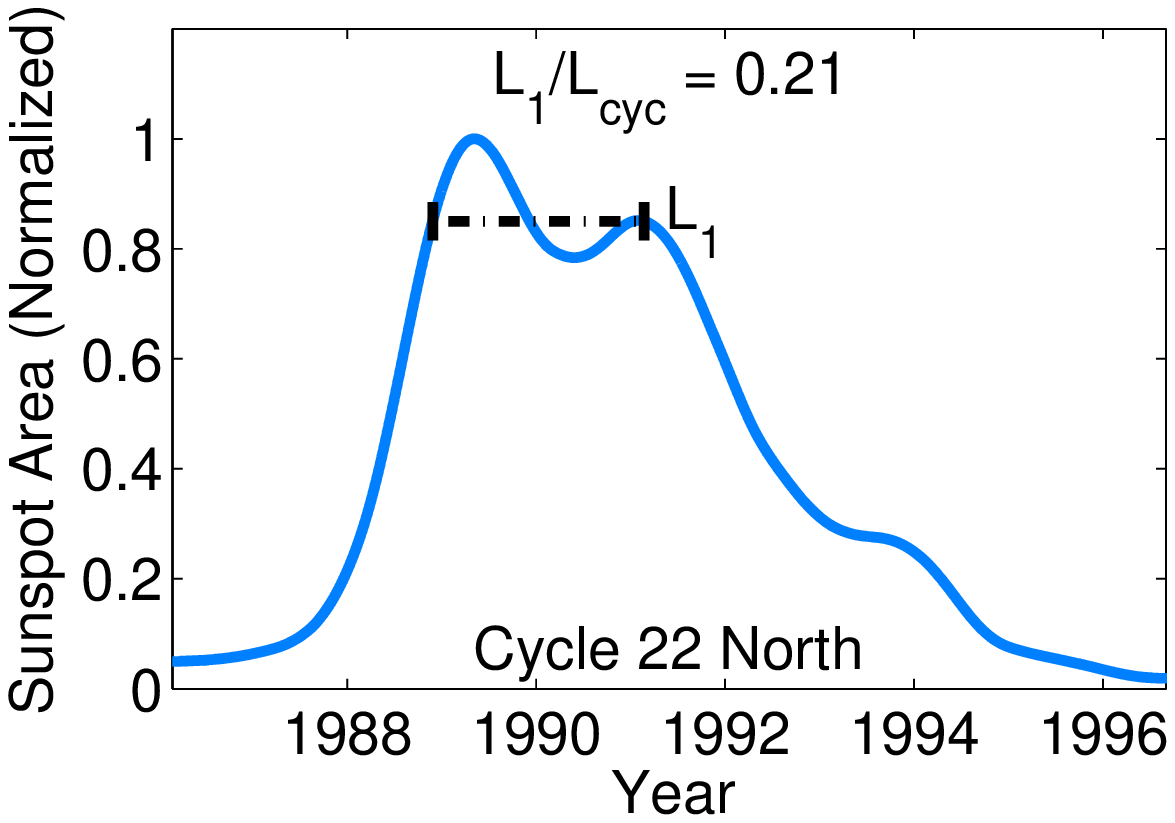}\\
  \includegraphics[width=0.32\textwidth]{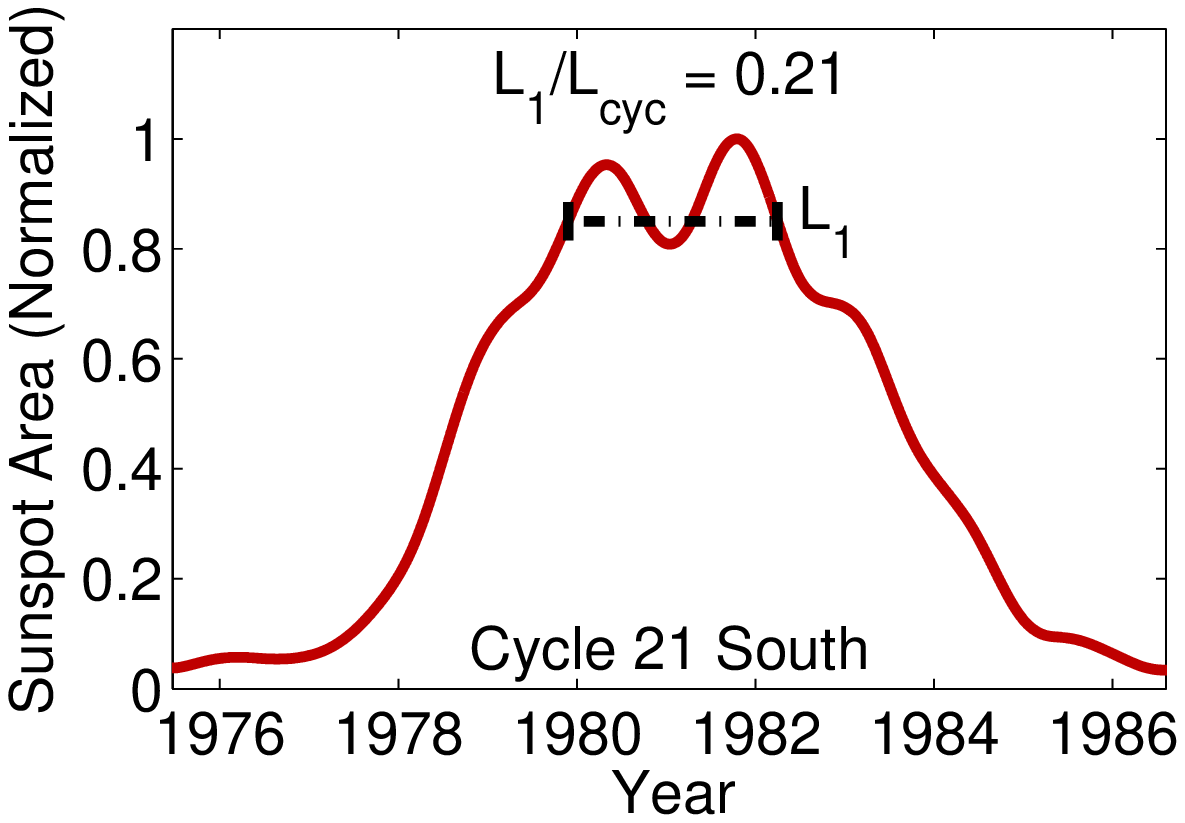} & \includegraphics[width=0.32\textwidth]{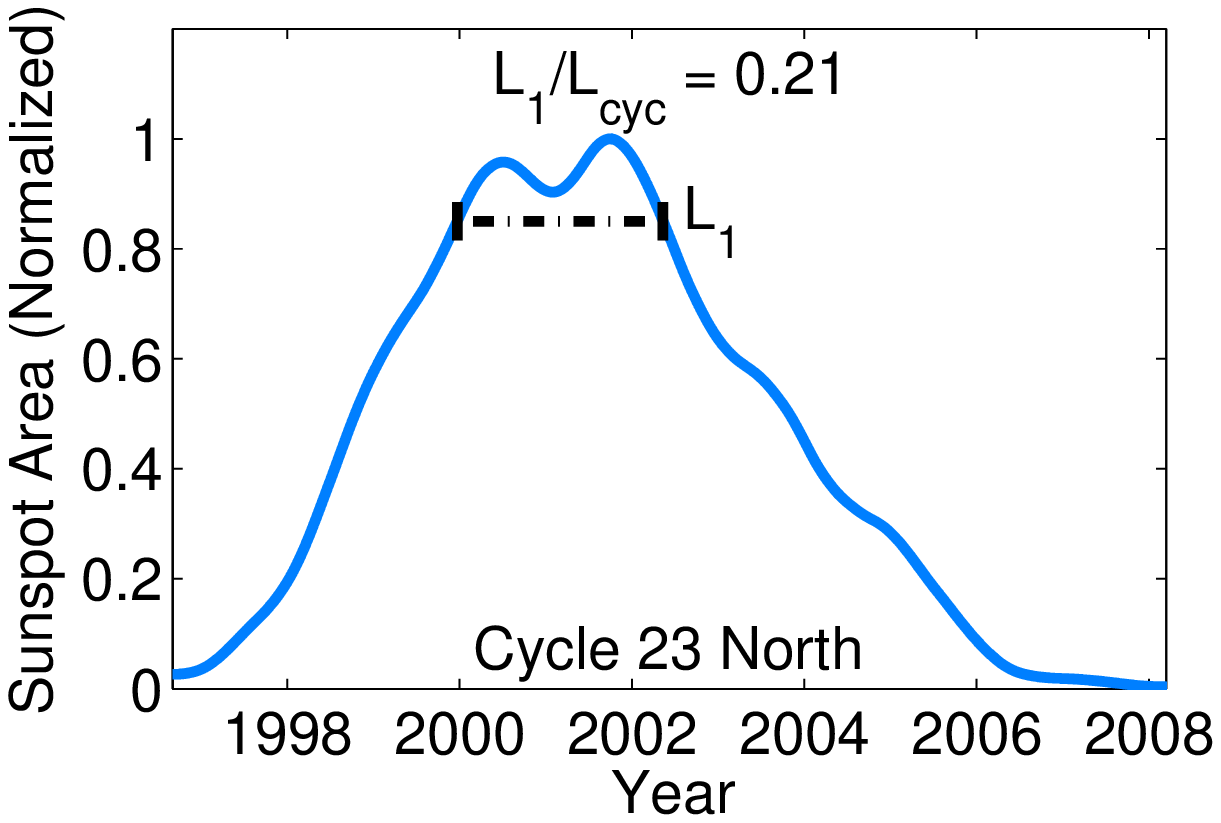} & \includegraphics[width=0.32\textwidth]{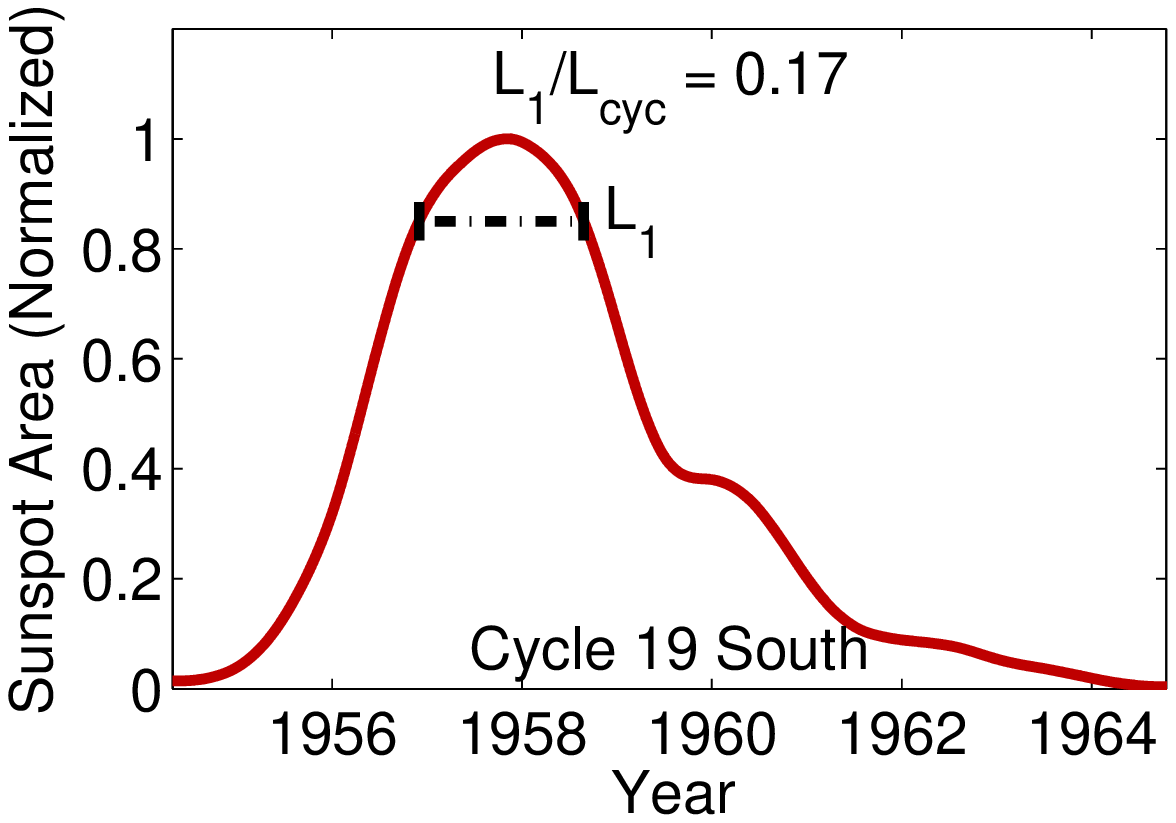}
\end{tabular}
\caption{Hemispheric cycles in descending order according to their WaM (Part 1).  Each cycle is normalized to its own maximum value and we calculate WaM using each cycle's width at 85\% of the total amplitude.}\label{Fig_Jggnss1}
\end{figure}

\begin{figure}
\centering
\begin{tabular}{ccc}
  \includegraphics[width=0.32\textwidth]{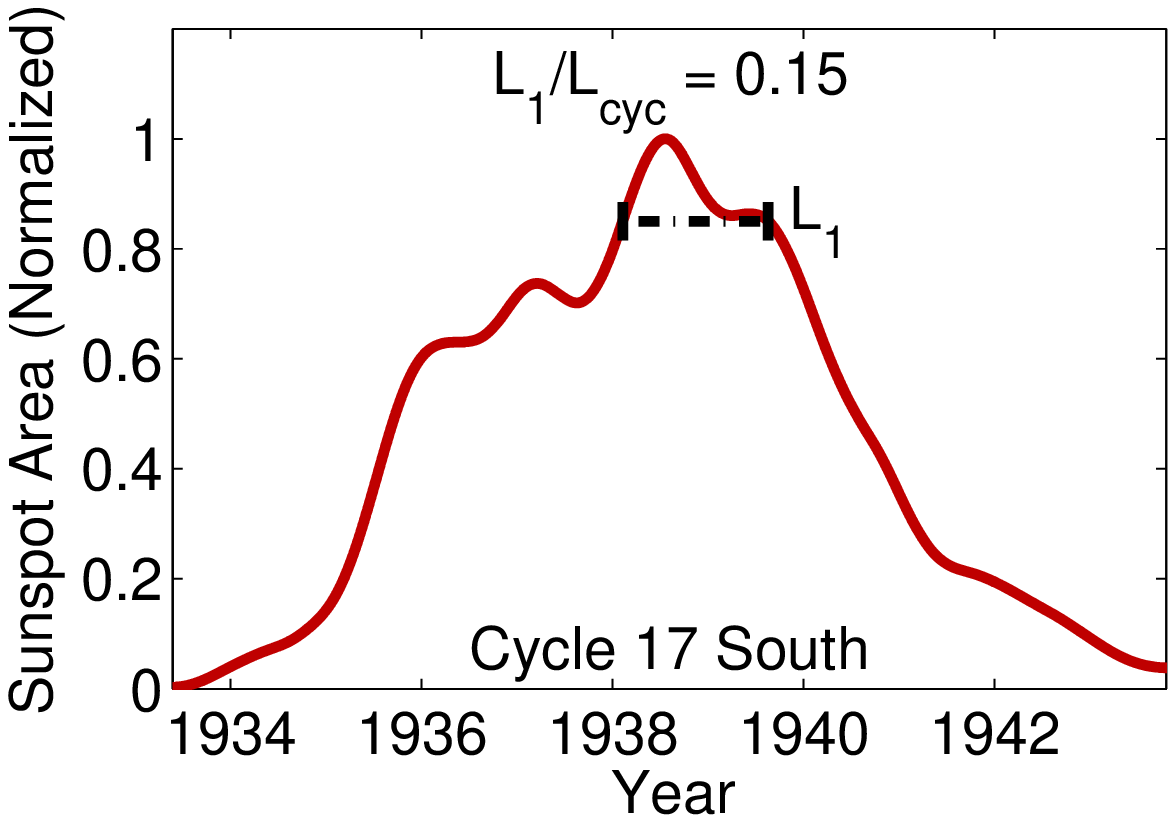} & \includegraphics[width=0.32\textwidth]{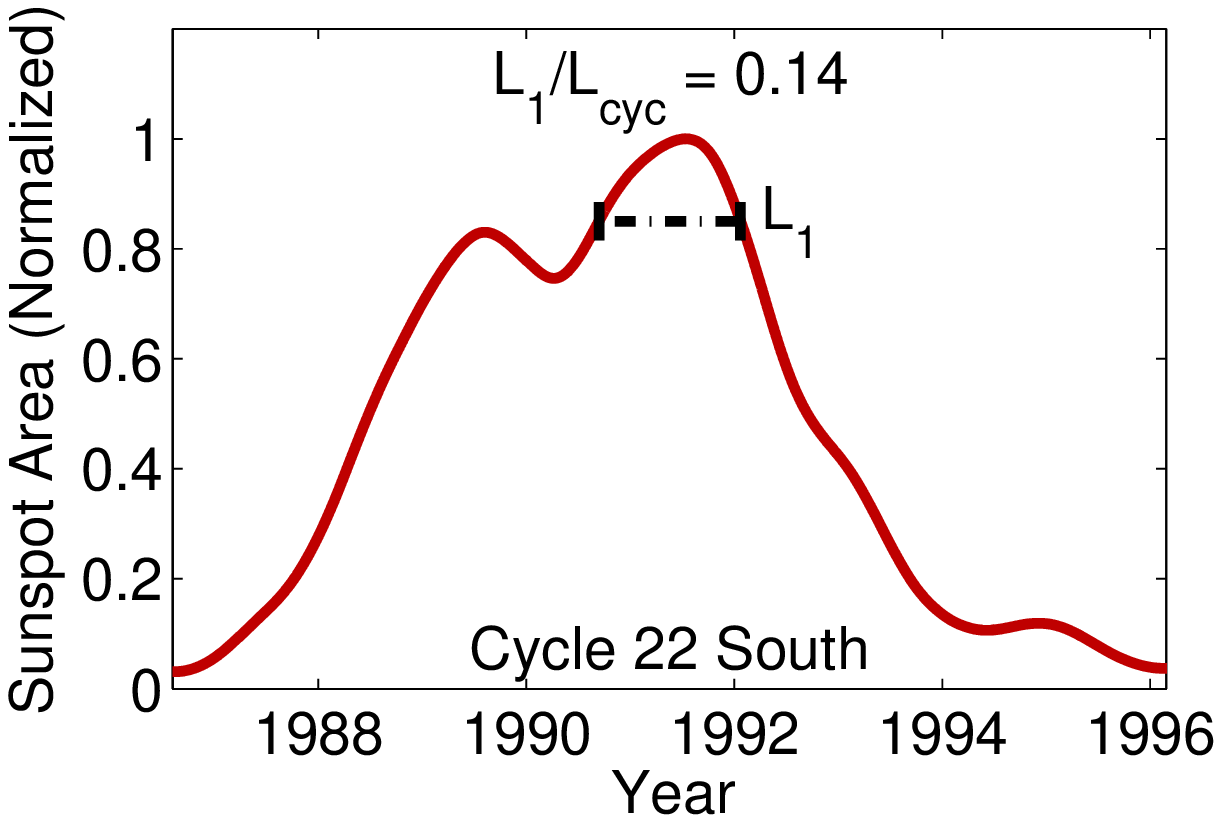} & \includegraphics[width=0.32\textwidth]{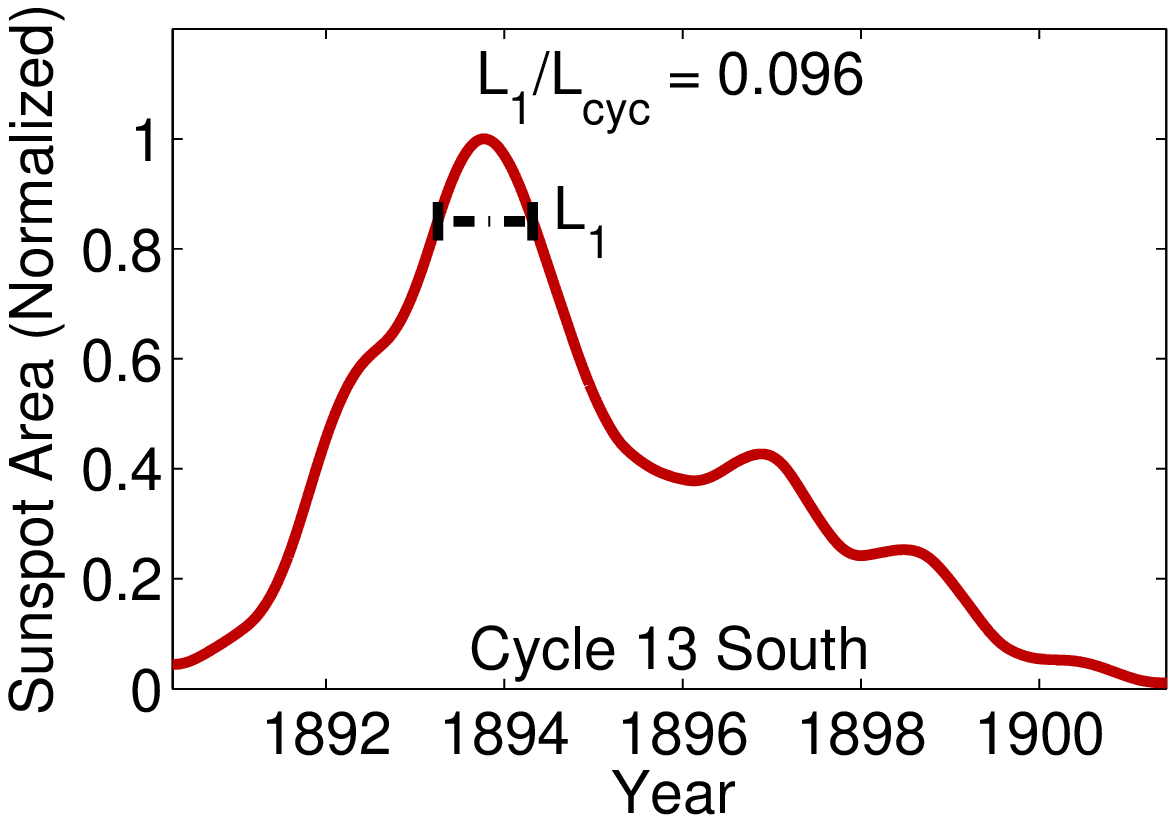}\\
  \includegraphics[width=0.32\textwidth]{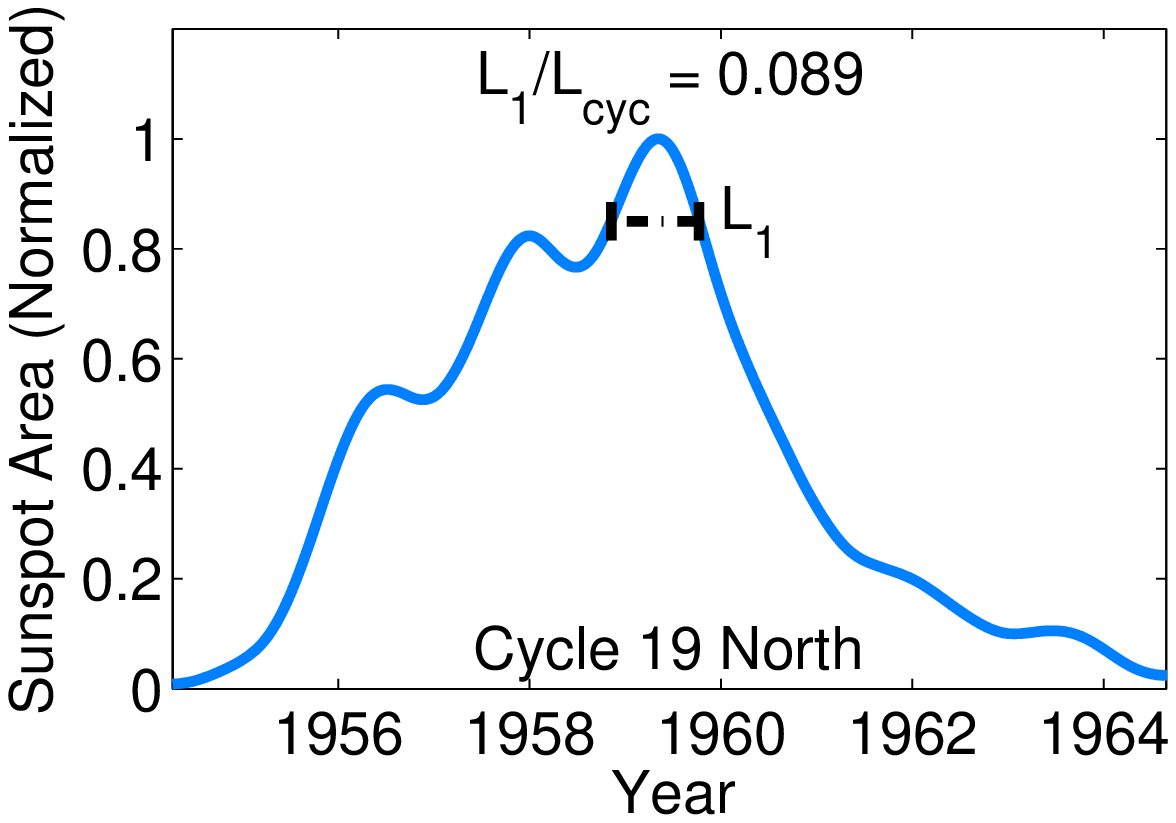} & \includegraphics[width=0.32\textwidth]{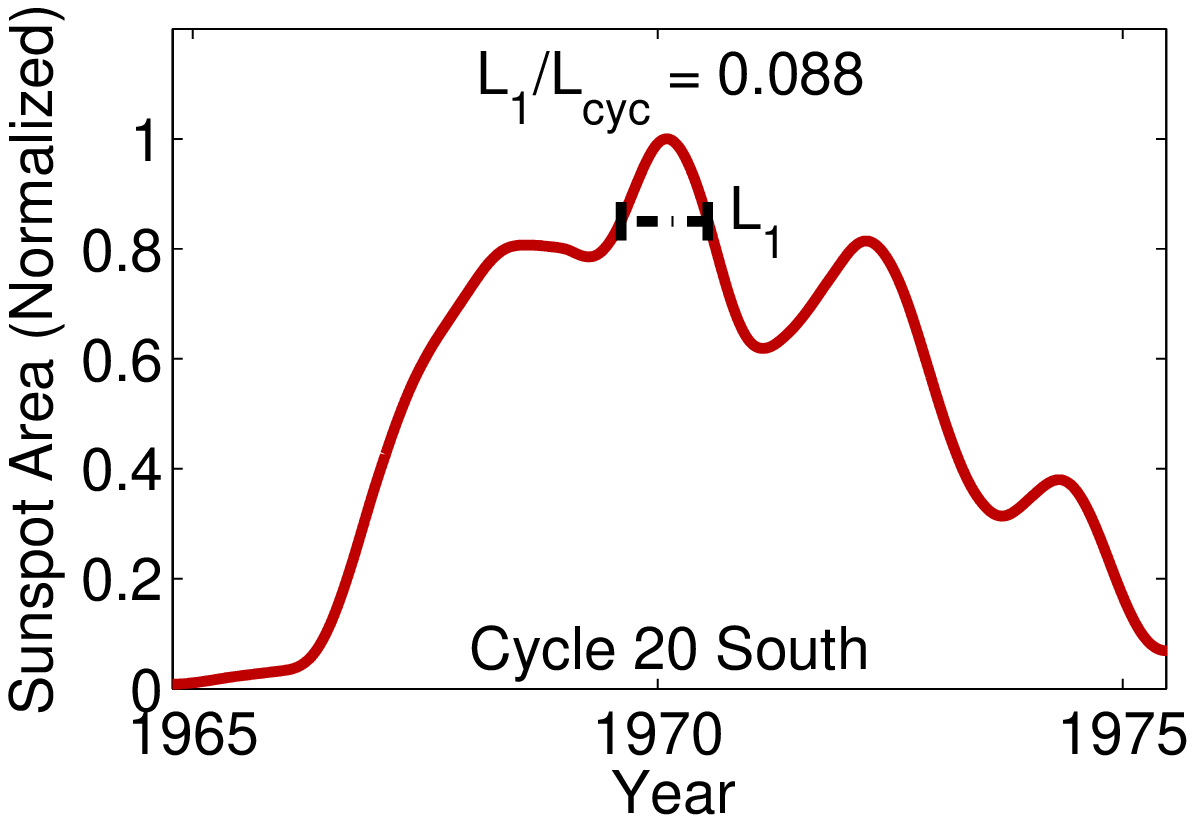} & \includegraphics[width=0.32\textwidth]{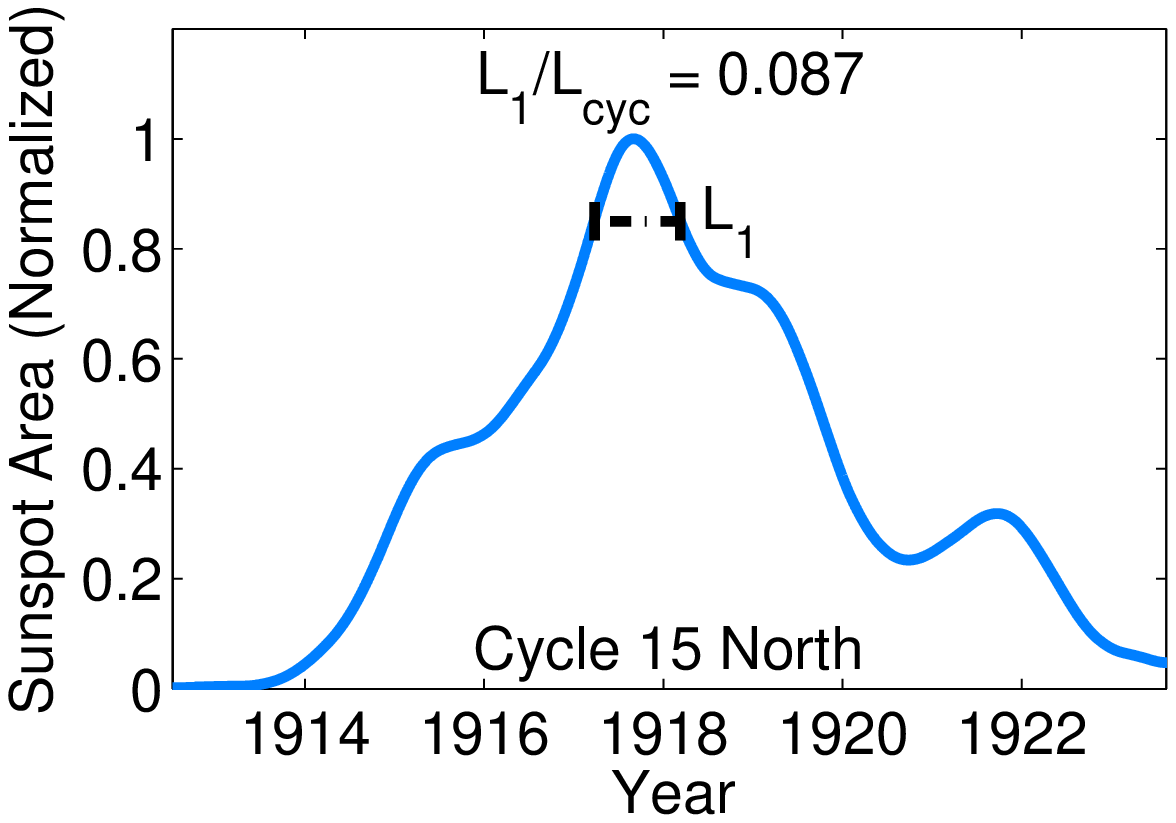}\\
  \includegraphics[width=0.32\textwidth]{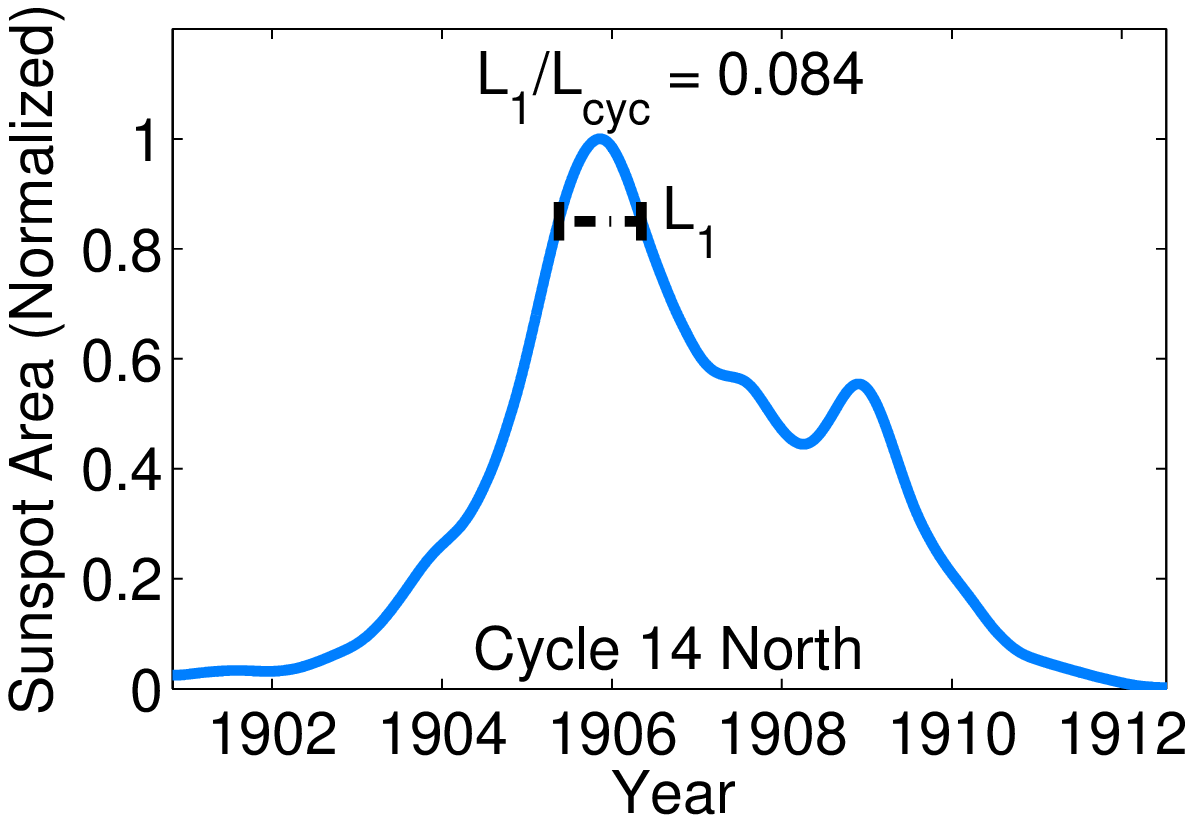} & \includegraphics[width=0.32\textwidth]{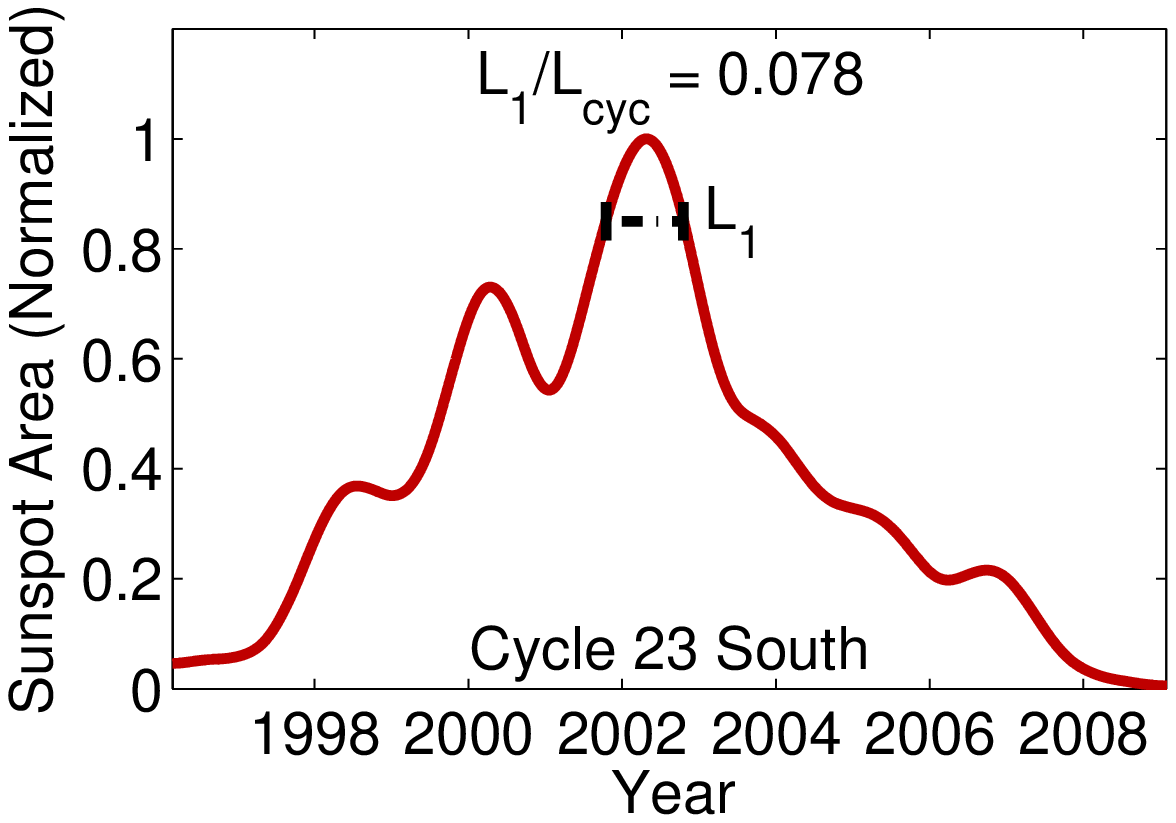} & \includegraphics[width=0.32\textwidth]{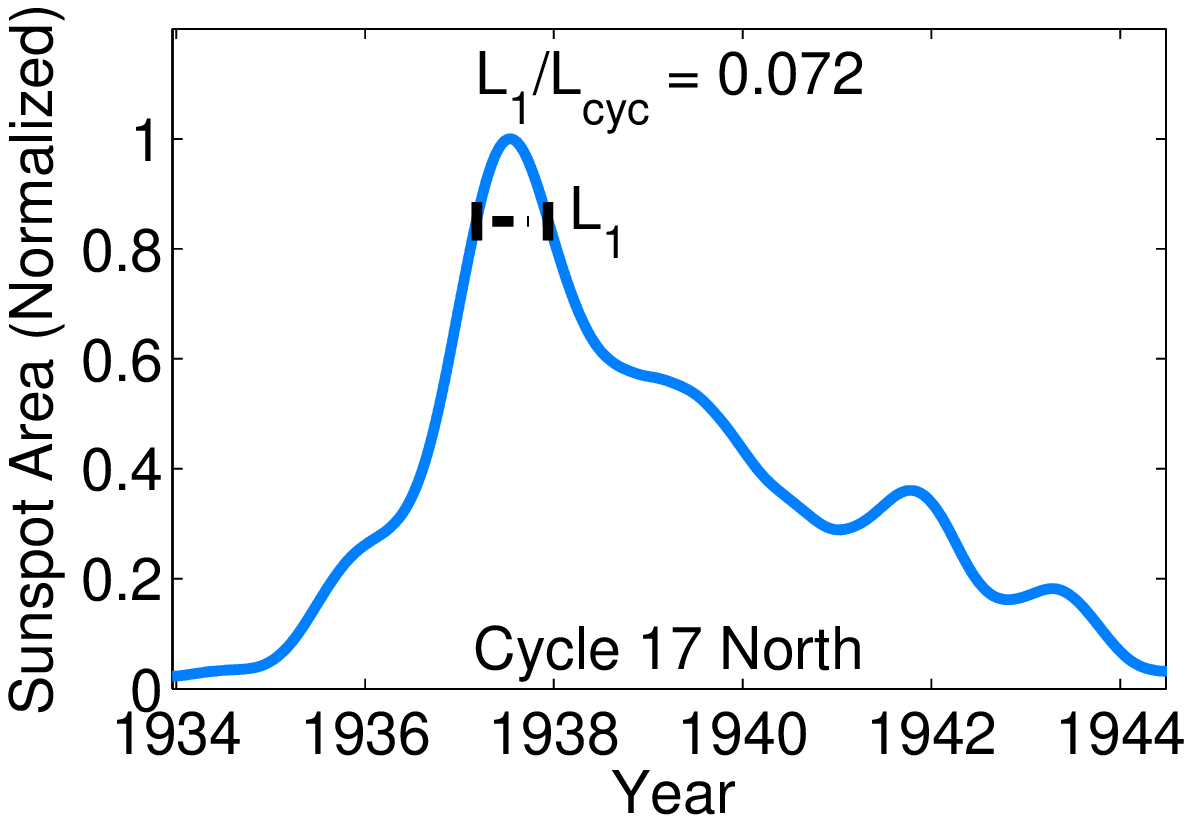}\\
  \includegraphics[width=0.32\textwidth]{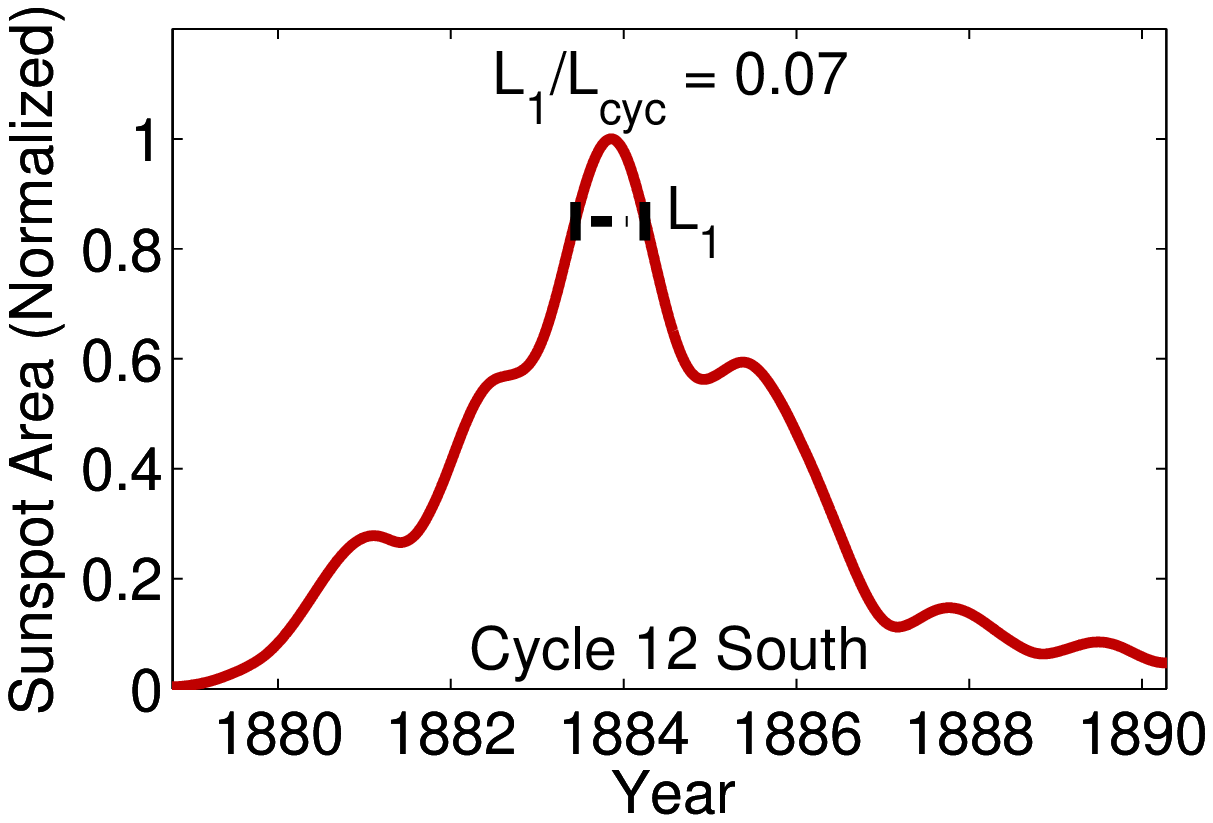} & \includegraphics[width=0.32\textwidth]{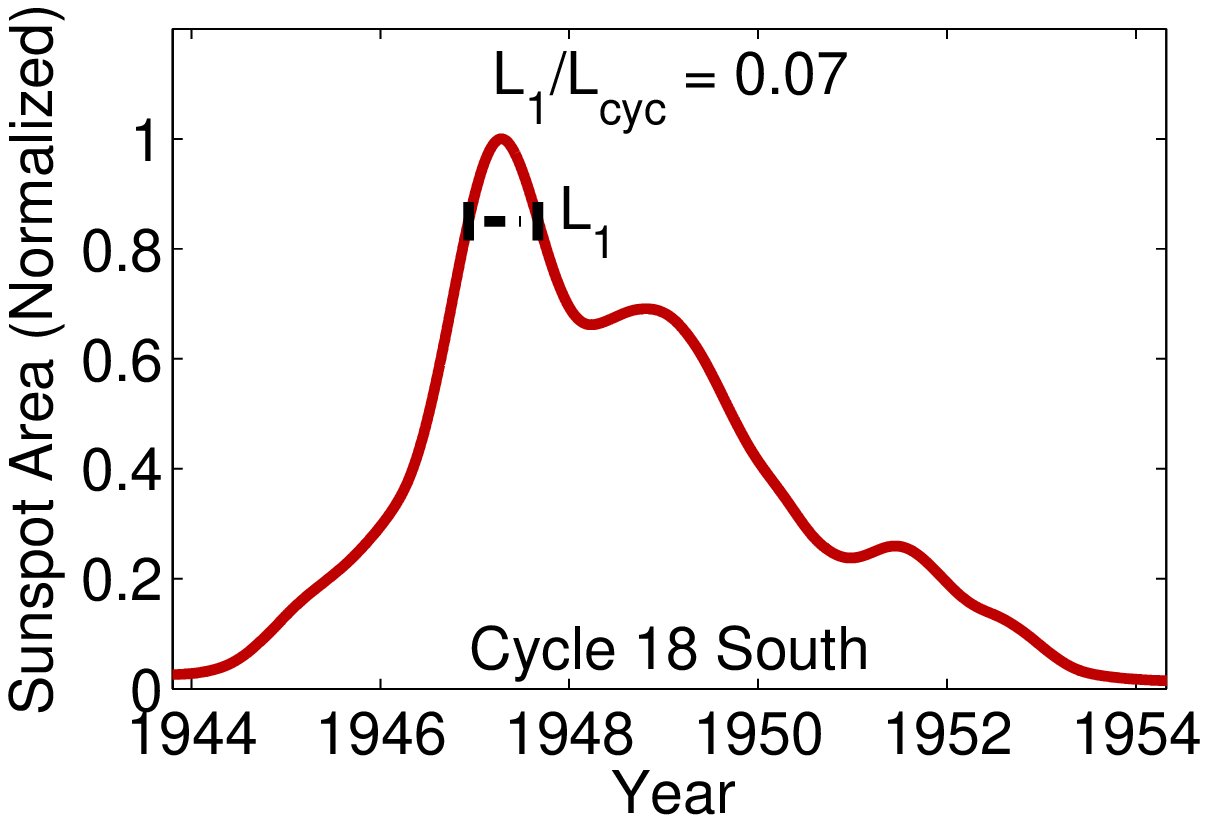} & \includegraphics[width=0.32\textwidth]{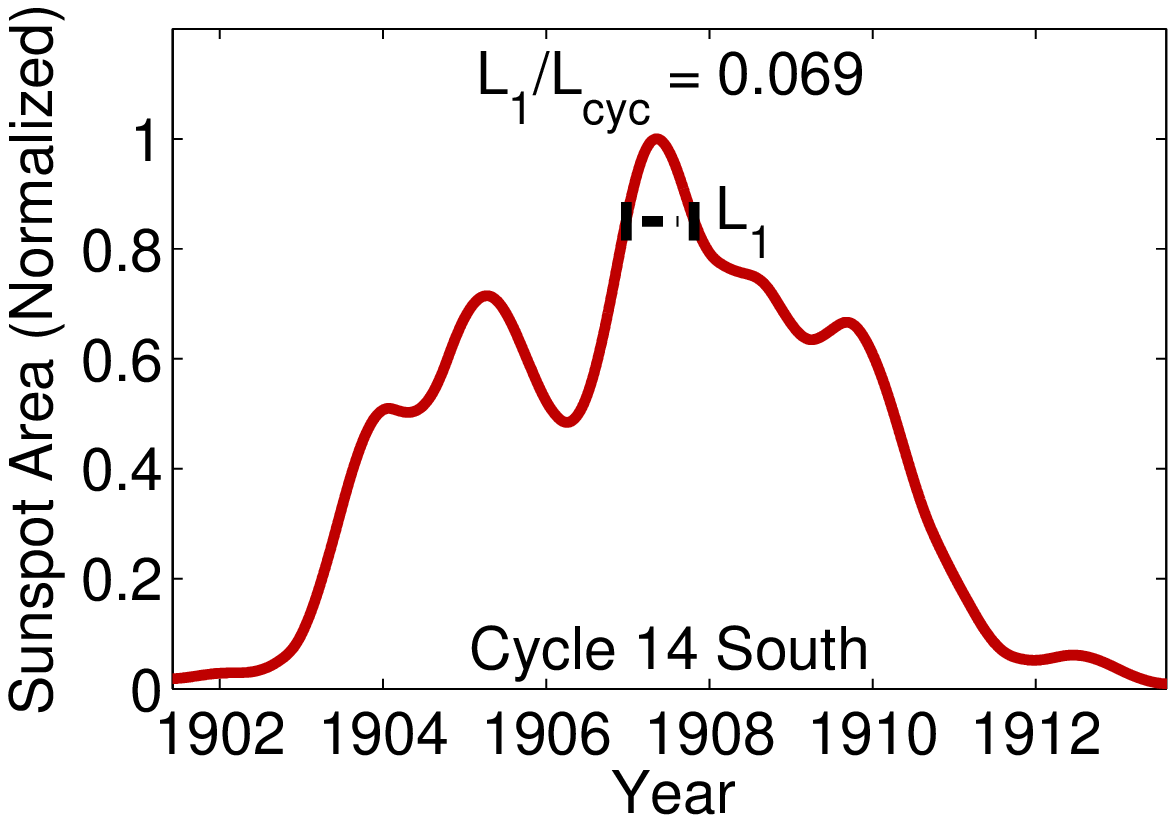}
\end{tabular}
\caption{Hemispheric cycles in descending order according to their WaM (Part 2). Each cycle is normalized to its own maximum value and we calculate WaM using each cycle's width at 85\% of the total amplitude.}\label{Fig_Jggnss2}
\end{figure}

While some cycles are characterized by a relatively flat and long-lasting maxima (or a sequence of peaks of similar amplitude, like cycle 18 north), most cycles are dominated by a single sharp and narrow peak (like cycle 18 south).  This characteristic can be quantified by calculating the ratio between their full width at maximum ($L_1$) and their total duration ($L_{cyc}$); from now on referred to as \emph{width at maximum} ($WaM = L_1/L_{cyc}$).  Figures \ref{Fig_Jggnss1} \& \ref{Fig_Jggnss2} show all cycles in our database arranged by WaM in decreasing order.

\begin{figure}[t!]
\begin{tabular}{ccc}
  \includegraphics[width=0.3\textwidth]{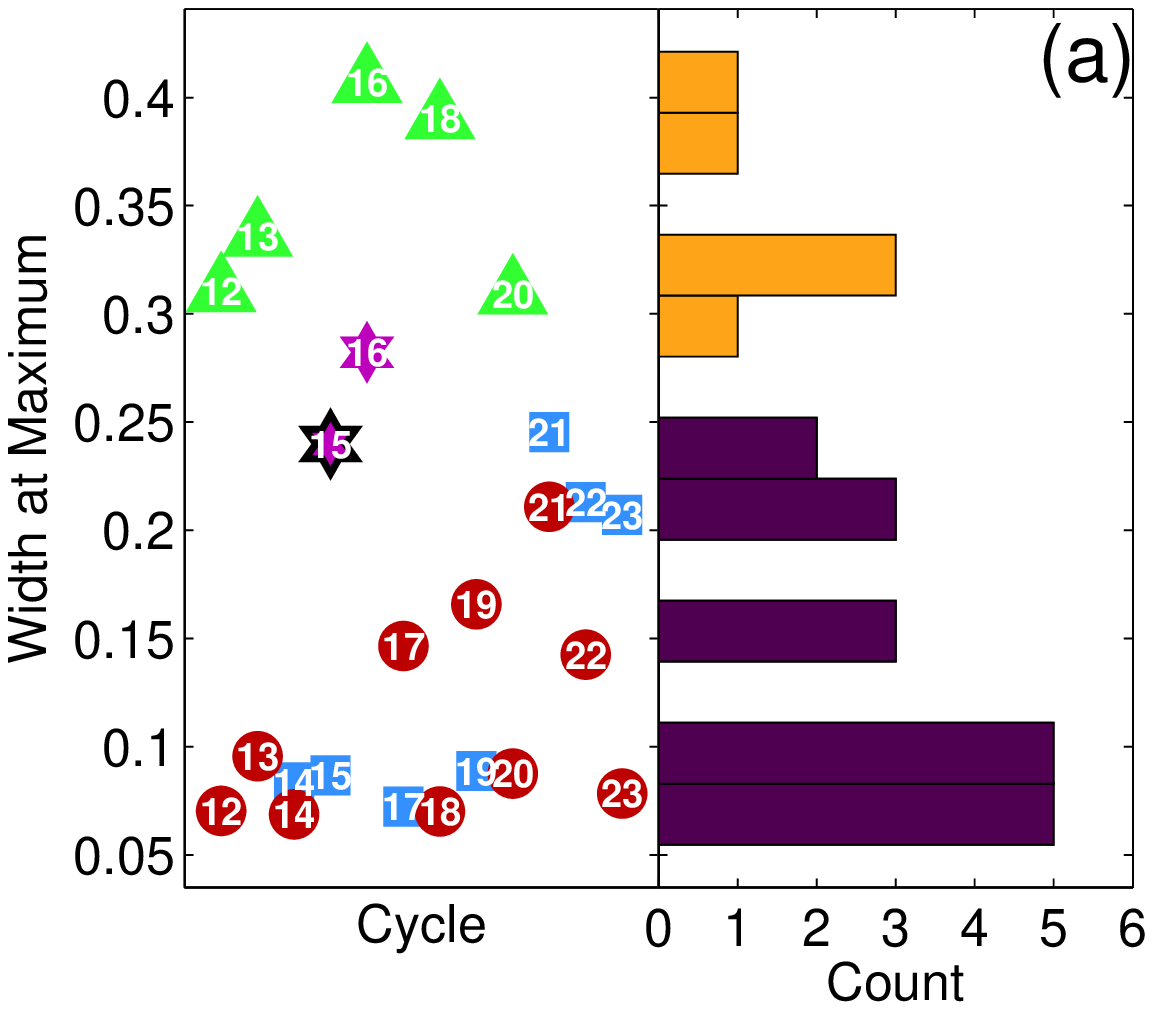} & \includegraphics[width=0.3\textwidth]{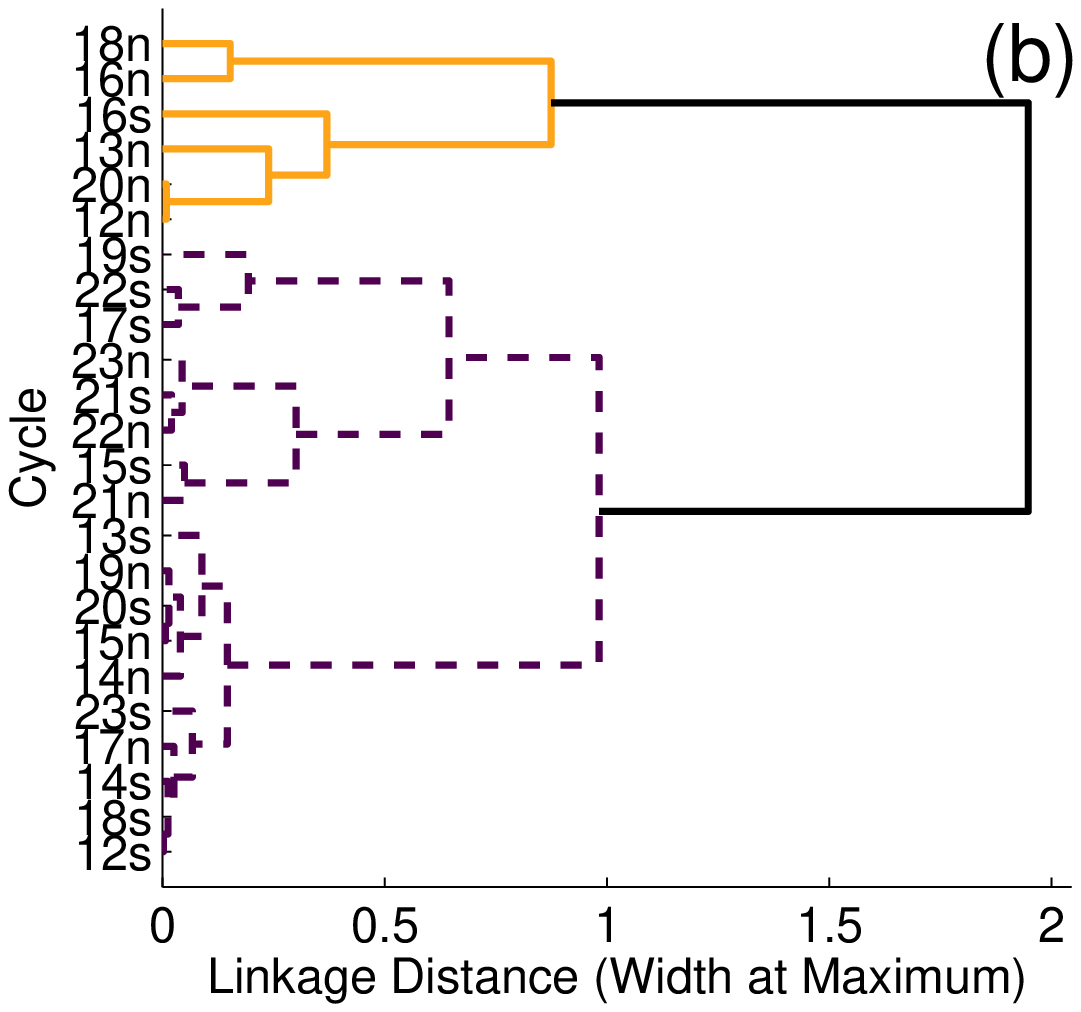} & \includegraphics[width=0.31\textwidth]{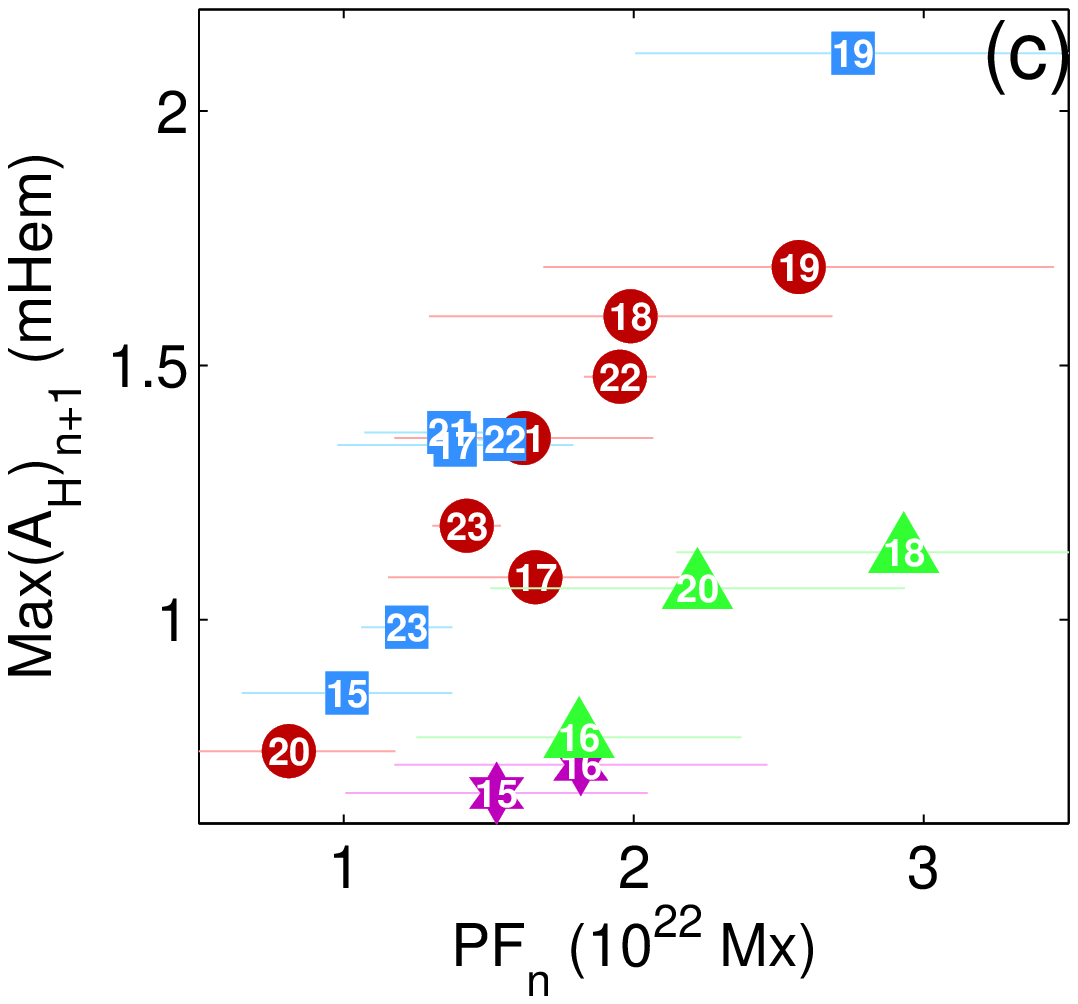}
\end{tabular}
\caption{Histogram (a) and dendogram (b) of cycle width at maximum.  Blue squares and green triangles (red circles and magenta stars) correspond to points in the Northern (Southern) hemisphere.  Natural clusters in width at maximum largely coincide with the branches in the relationship between polar flux at minimum vs.\ the amplitude of the next cycle (c).  The exception (highlighted in black in the histogram) is cycle 15 south. Note that cycles 12, 13 and 14 are missing because they are missing in our polar flux dataset (our sunspot area dataset spans a longer period of time).}\label{Fig_Jgdd}
\end{figure}

Looking at a histogram of WaM (Figure~\ref{Fig_Jgdd}-a) we find that cycles seem to be grouped into separate clusters. We perform a hierarchical cluster analysis \cite{jain-dubes1988} in order to find out if this represents a natural division in the data.  The basic idea behind cluster analysis is to classify a set of data into separate groups (clusters) such that elements within a cluster are more alike than elements belonging to separate clusters.  In particular, we are interested in using WaM as our clustering criteria and define cycles to be more similar the closer their WaM values are.  Hierarchical cluster analysis starts by pairing points based on their proximity to form the first level of clusters and progresses by pairing nearby clusters (forming bigger and bigger groups) until all points are linked in the hierarchy.  This process can be nicely visualized using a hierarchical diagram (commonly referred to as a dendrogram).  Links between objects are represented by U-shaped lines and the arm-length of the U indicates the distance between the objects; a natural division in the data occurs when the length of a link differs noticeably from the length of the neighboring links below it (this is referred to as link inconsistency).

Figure \ref{Fig_Jgdd}-b shows the dendrogram obtained using cycle WaM.  We calculate the similarity between cycles using the Euclidean distance and perform our linkage using the weighted average distance.  This gives us the best cophenetic correlation coefficient\cite{sokal-rohlf1962} ($c=0.82$), which is a measure of how faithfully the distances between two points are preserved in the dendrogram (the closer to $1$ the better).  As can be observed in Figure~\ref{Fig_Jgdd}-b, WaM naturally separates cycles into two distinct clusters (the length of the last link is larger than the length of the links below it).   These clusters largely coincide with the two separate branches in the plot of polar flux at minimum vs.\ amplitude of the next cycle (Figure~\ref{Fig_Jgdd}-c); the one exception is the hemispheric cycle 15 south.

\begin{figure}
\centering
\begin{tabular}{ccc}
  \includegraphics[scale=0.4]{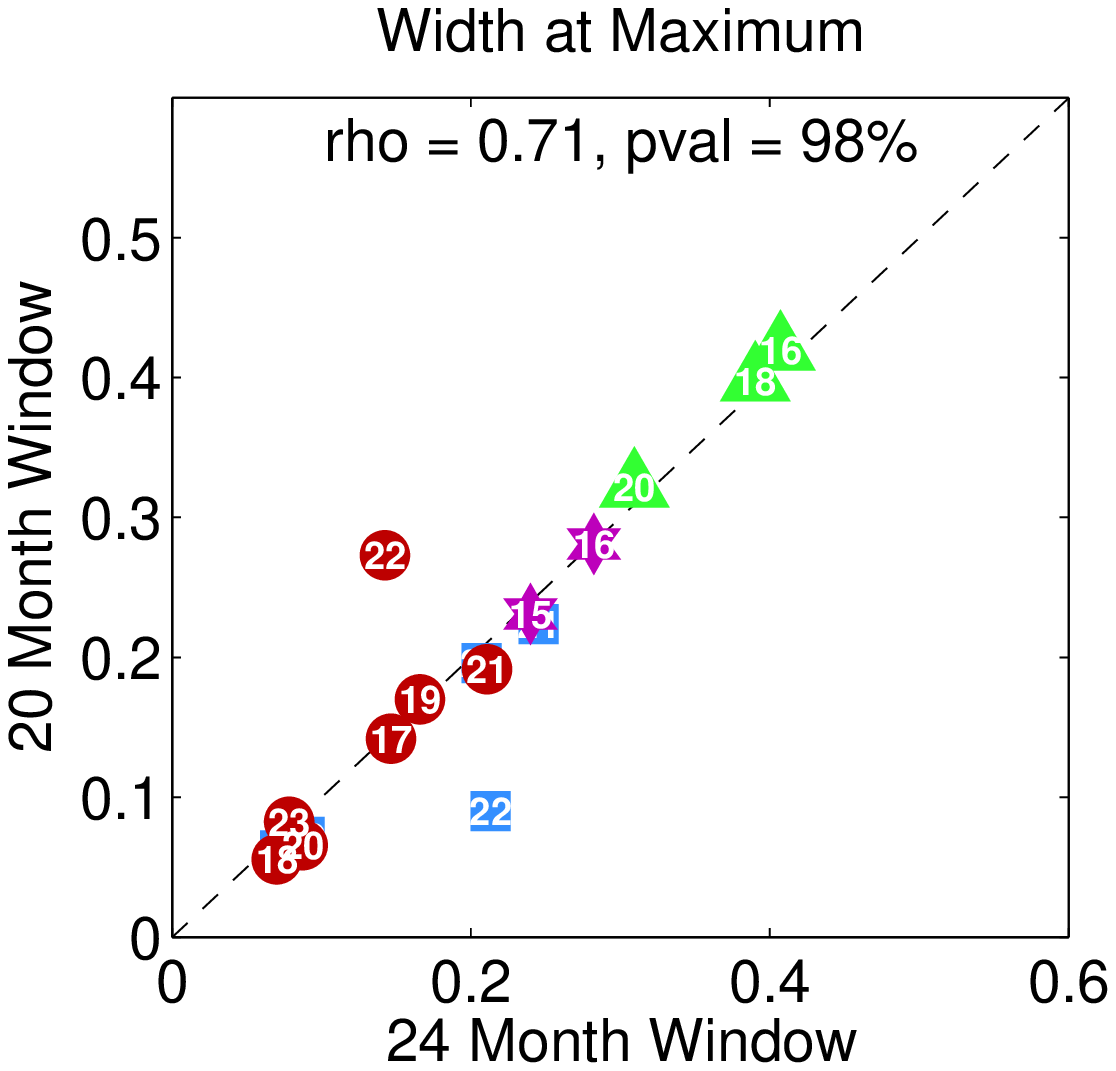} & \includegraphics[scale=0.4]{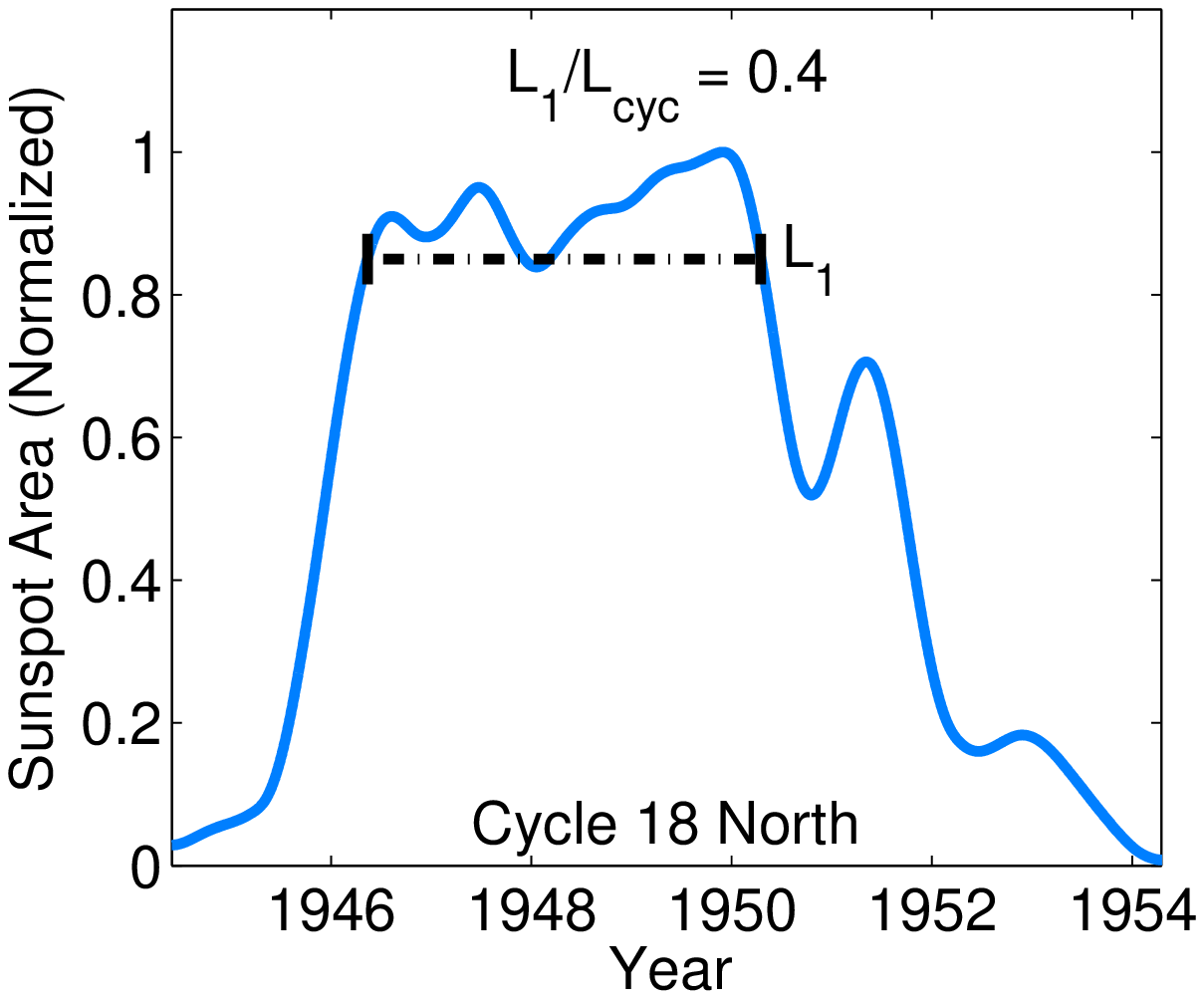} & \includegraphics[scale=0.4]{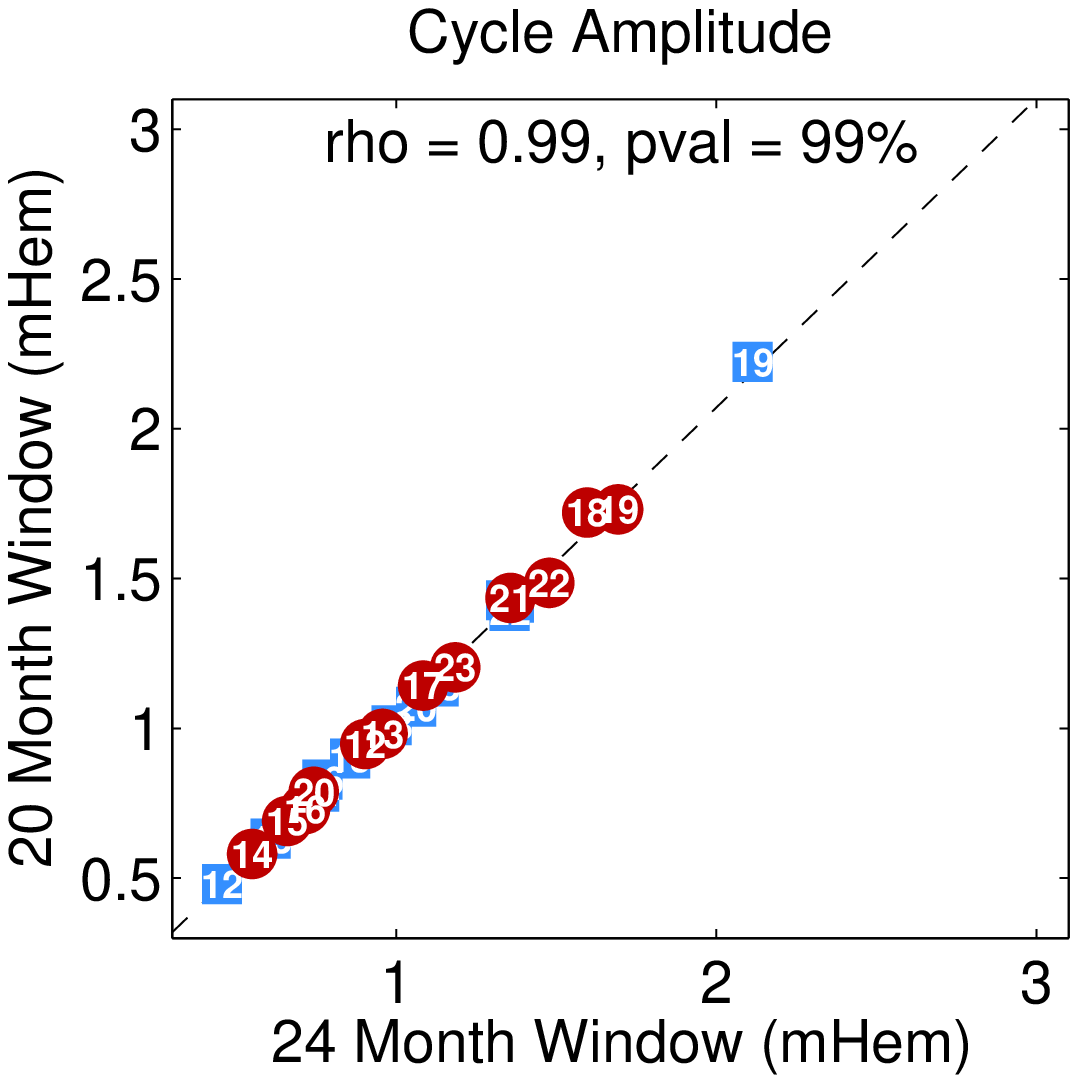}\\
  \includegraphics[scale=0.4]{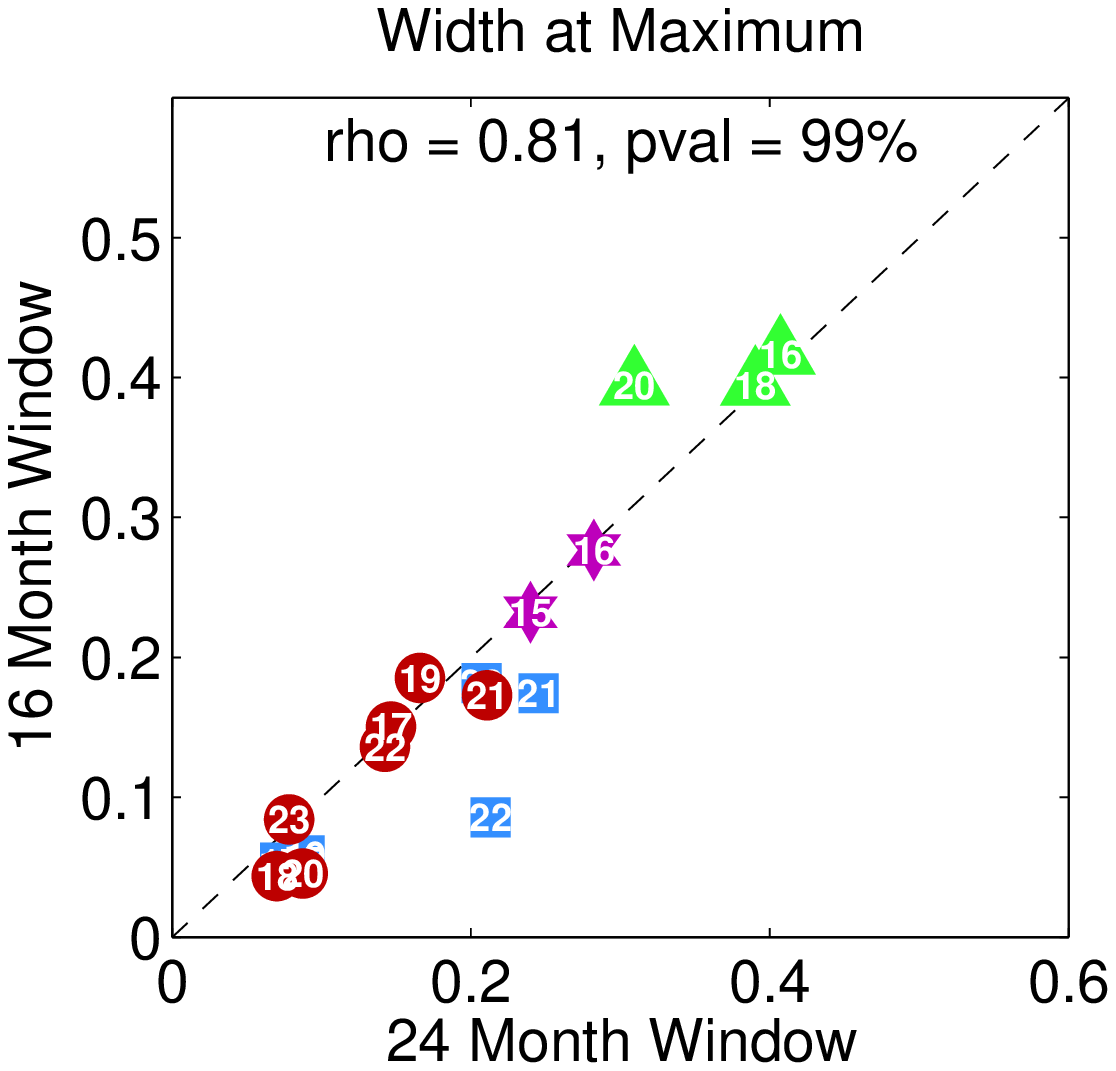} & \includegraphics[scale=0.4]{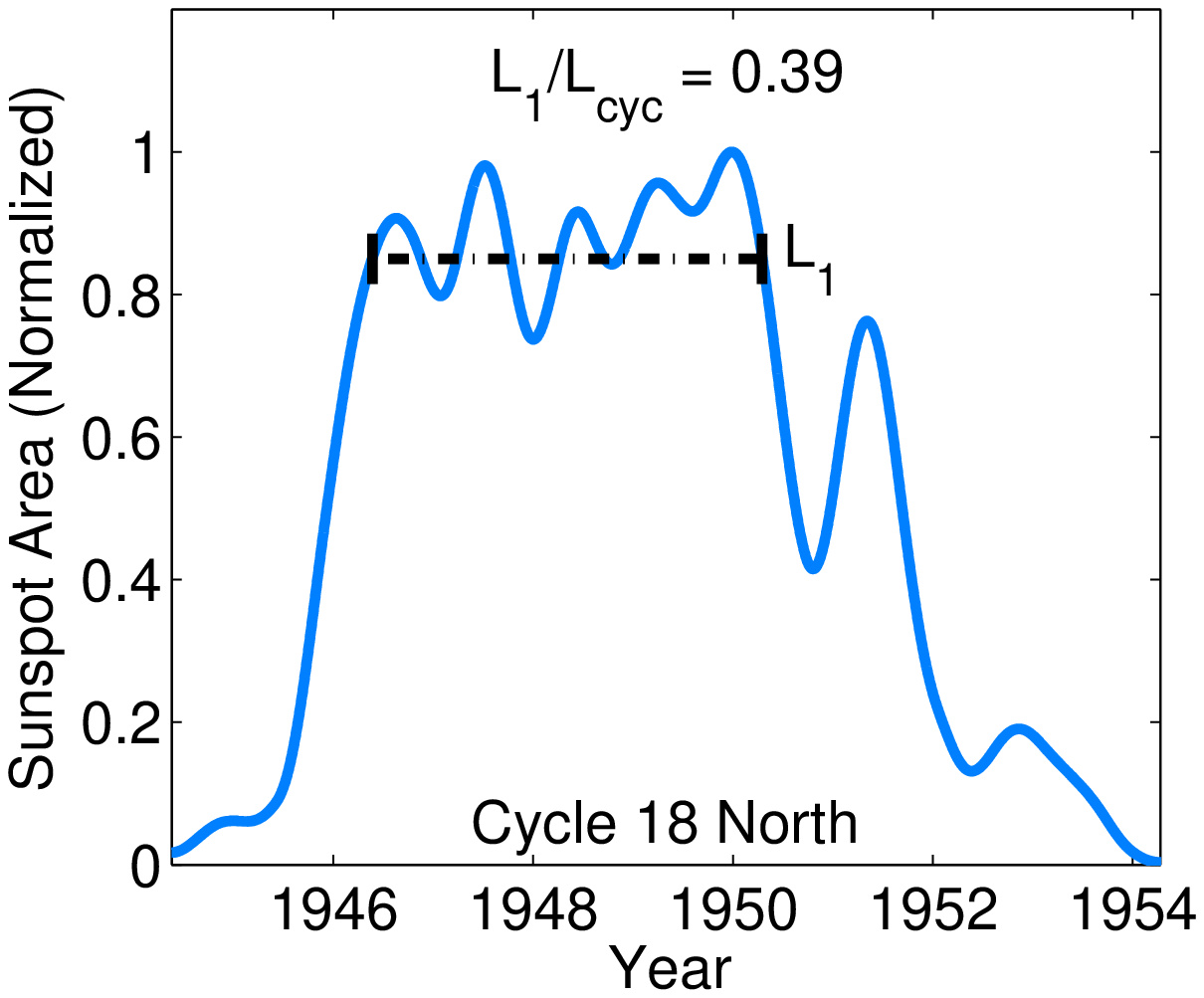} & \includegraphics[scale=0.4]{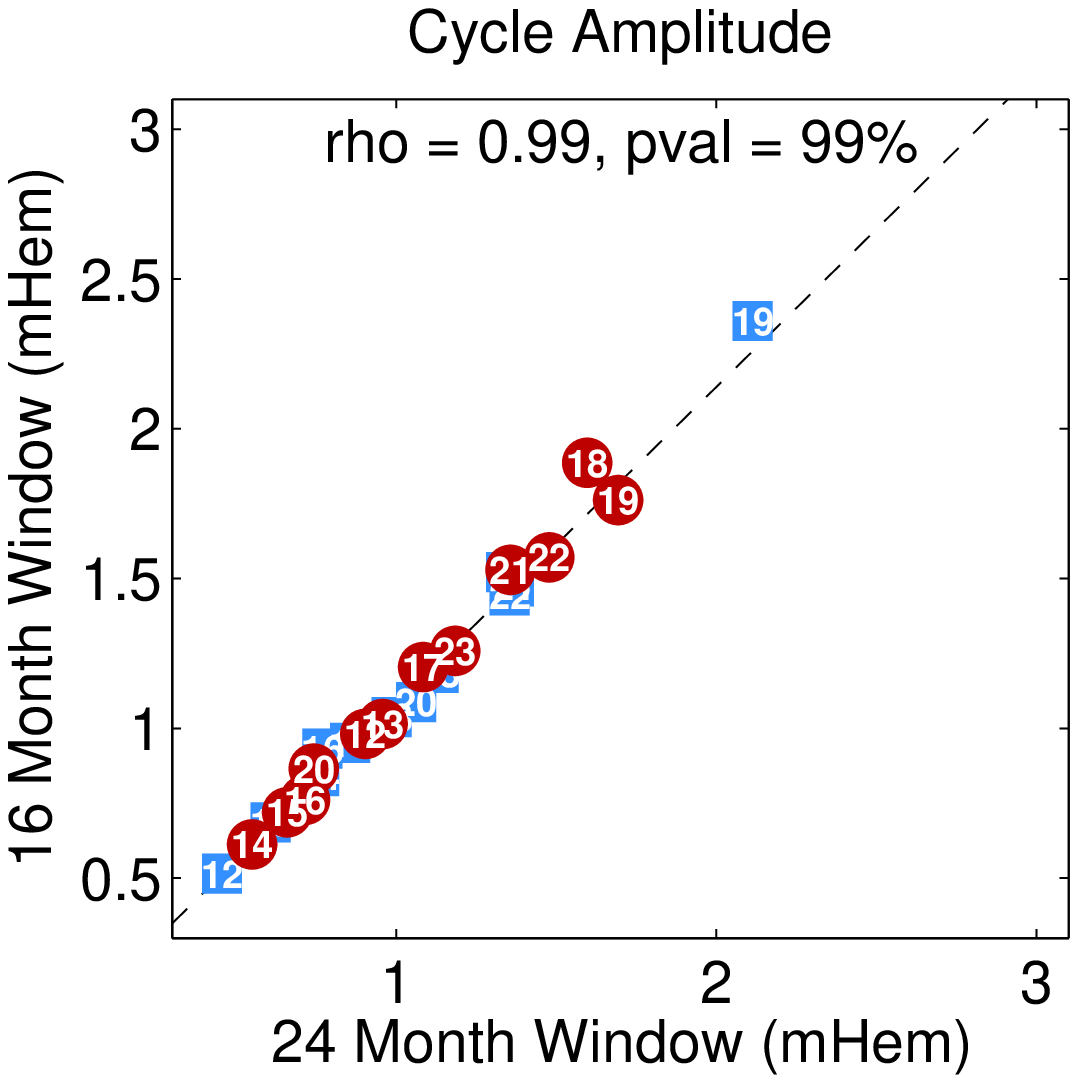}\\
  \includegraphics[scale=0.4]{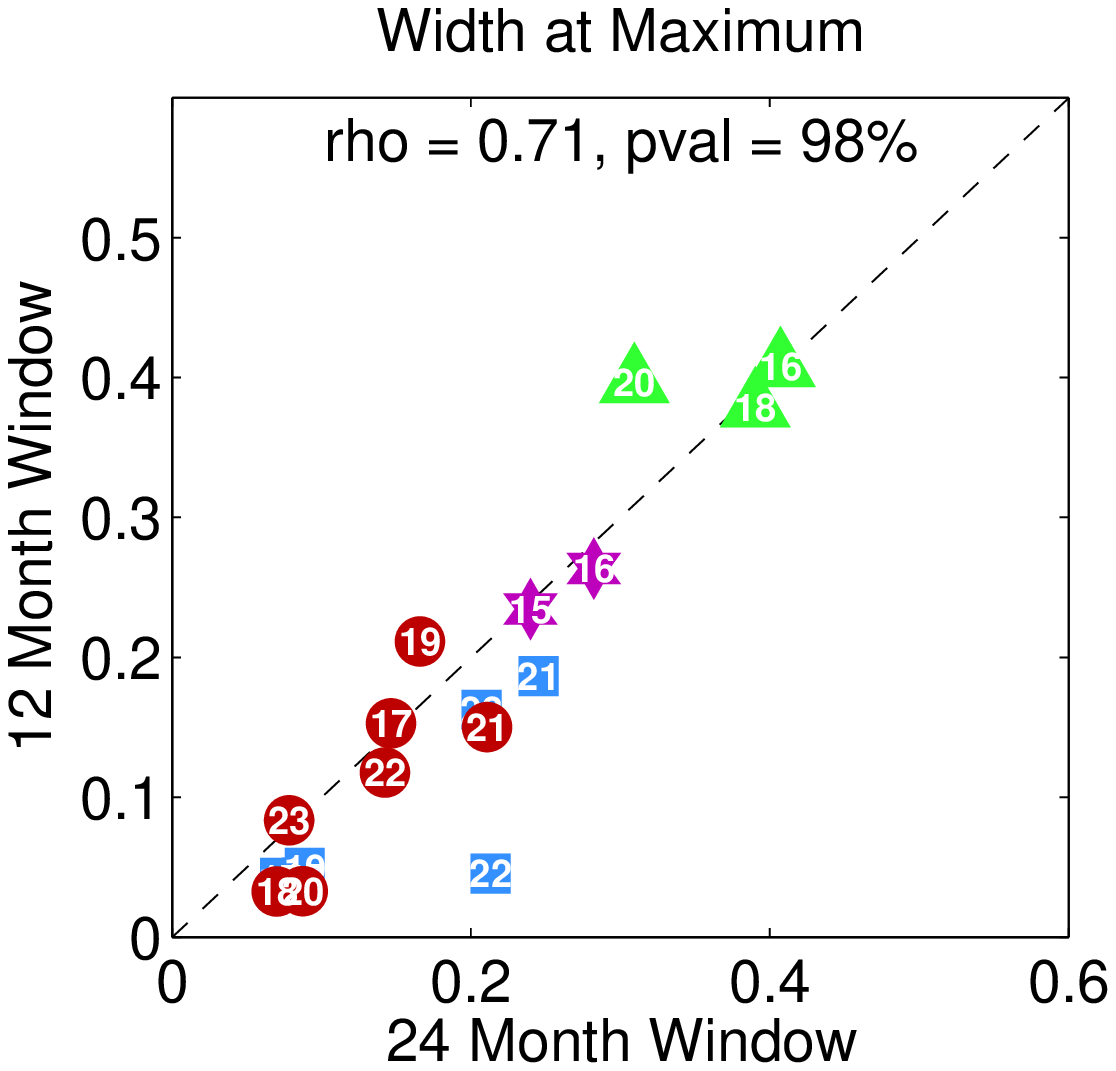} & \includegraphics[scale=0.4]{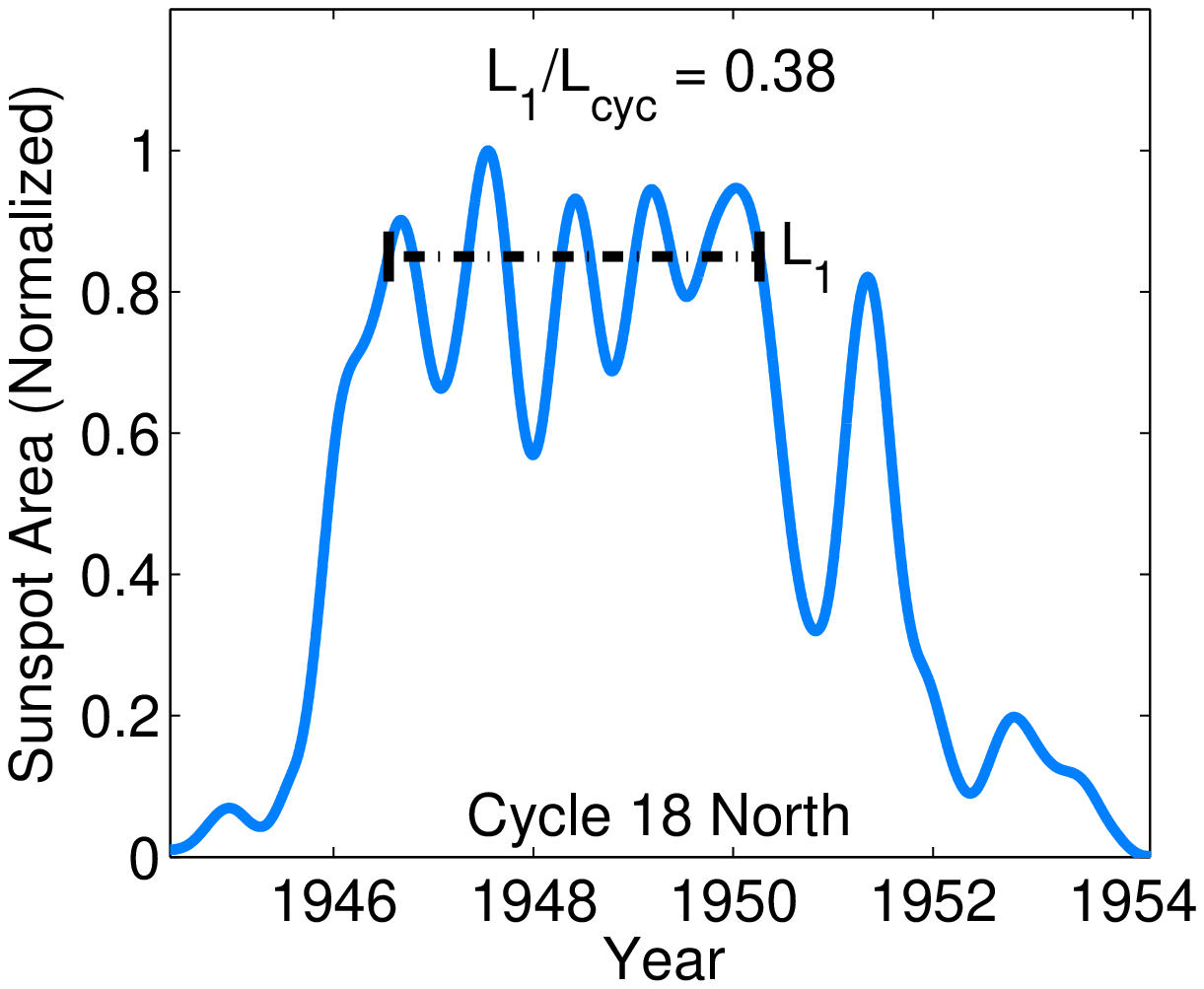} & \includegraphics[scale=0.4]{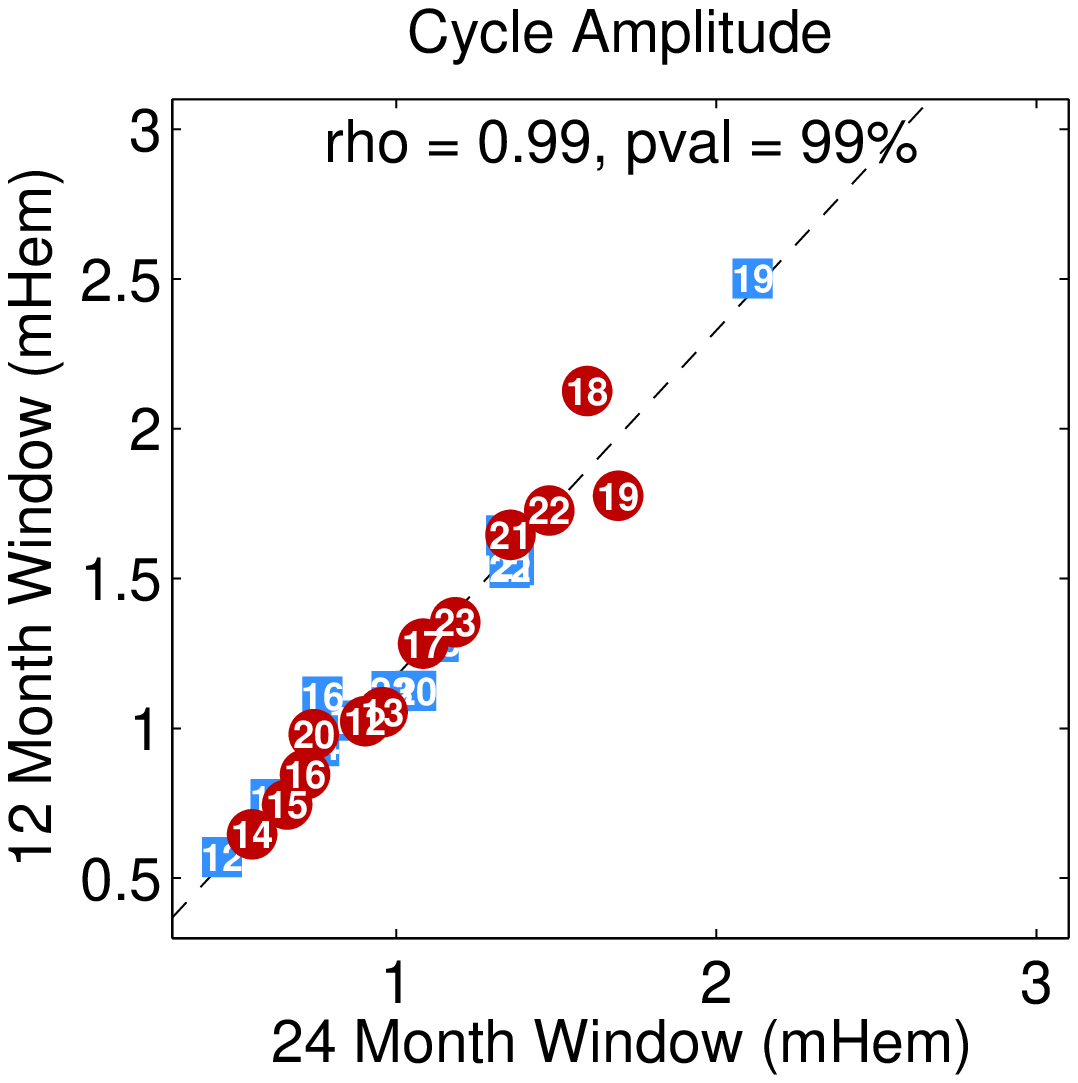}\\
  \includegraphics[scale=0.4]{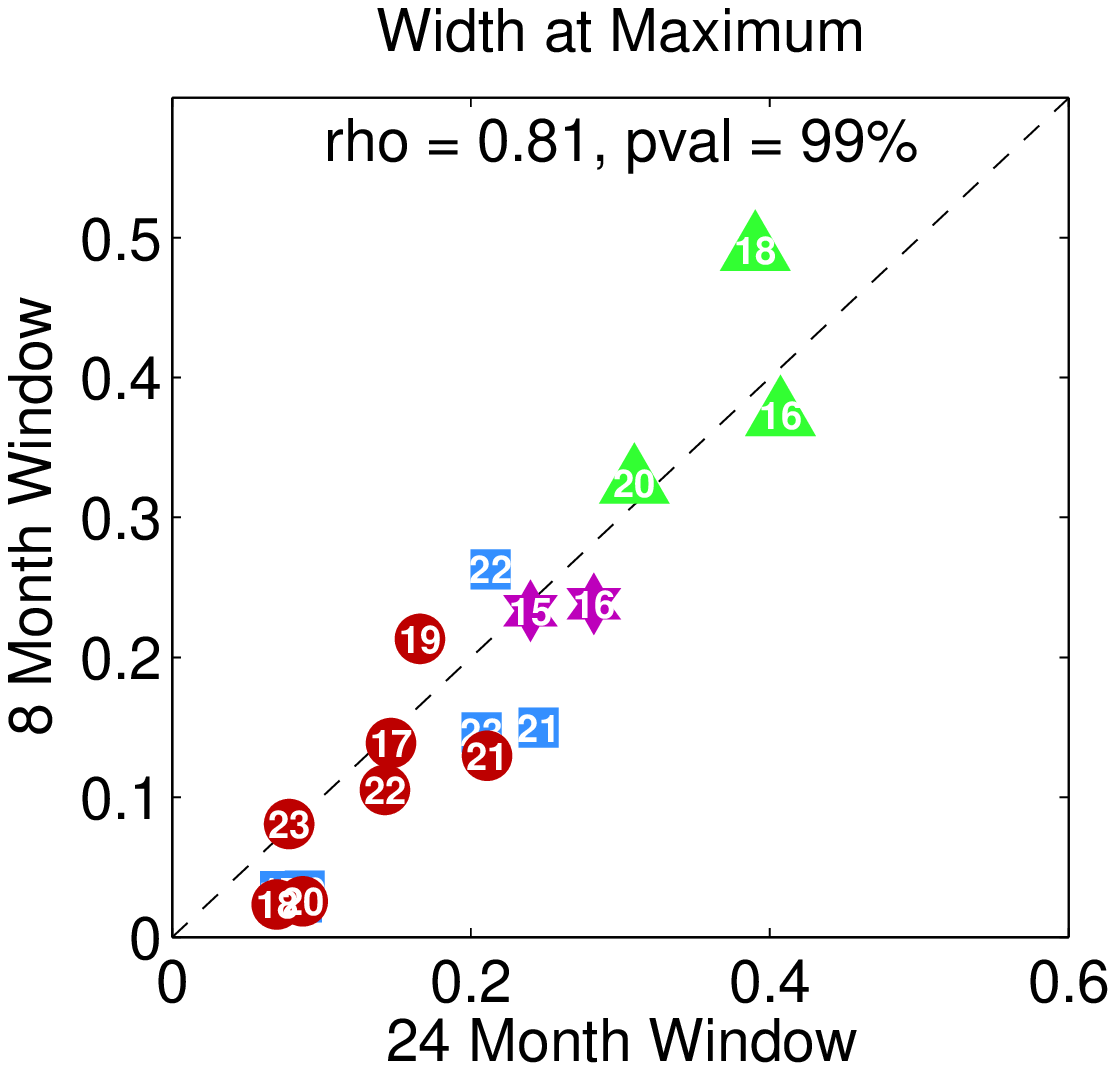} & \includegraphics[scale=0.4]{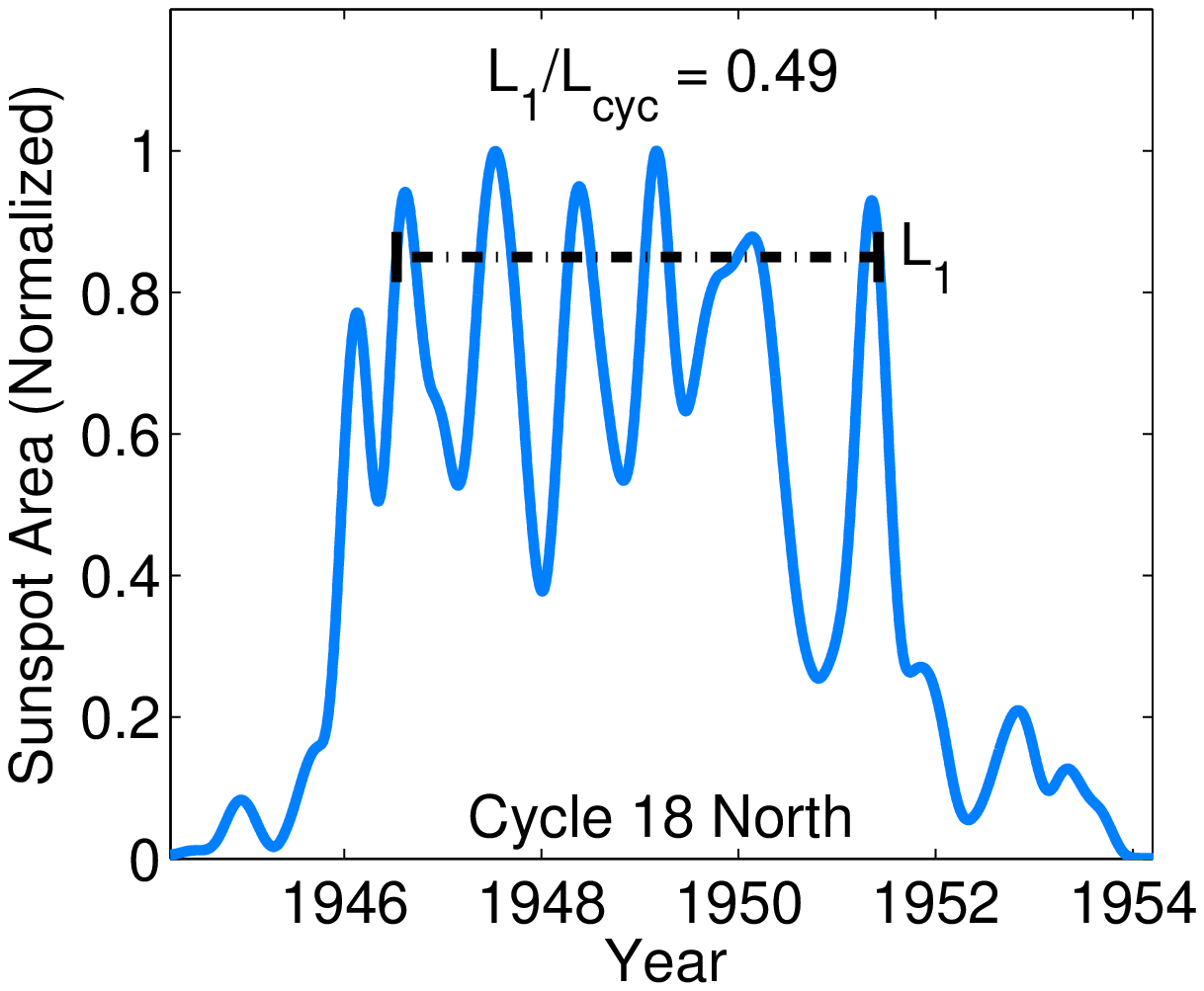} & \includegraphics[scale=0.4]{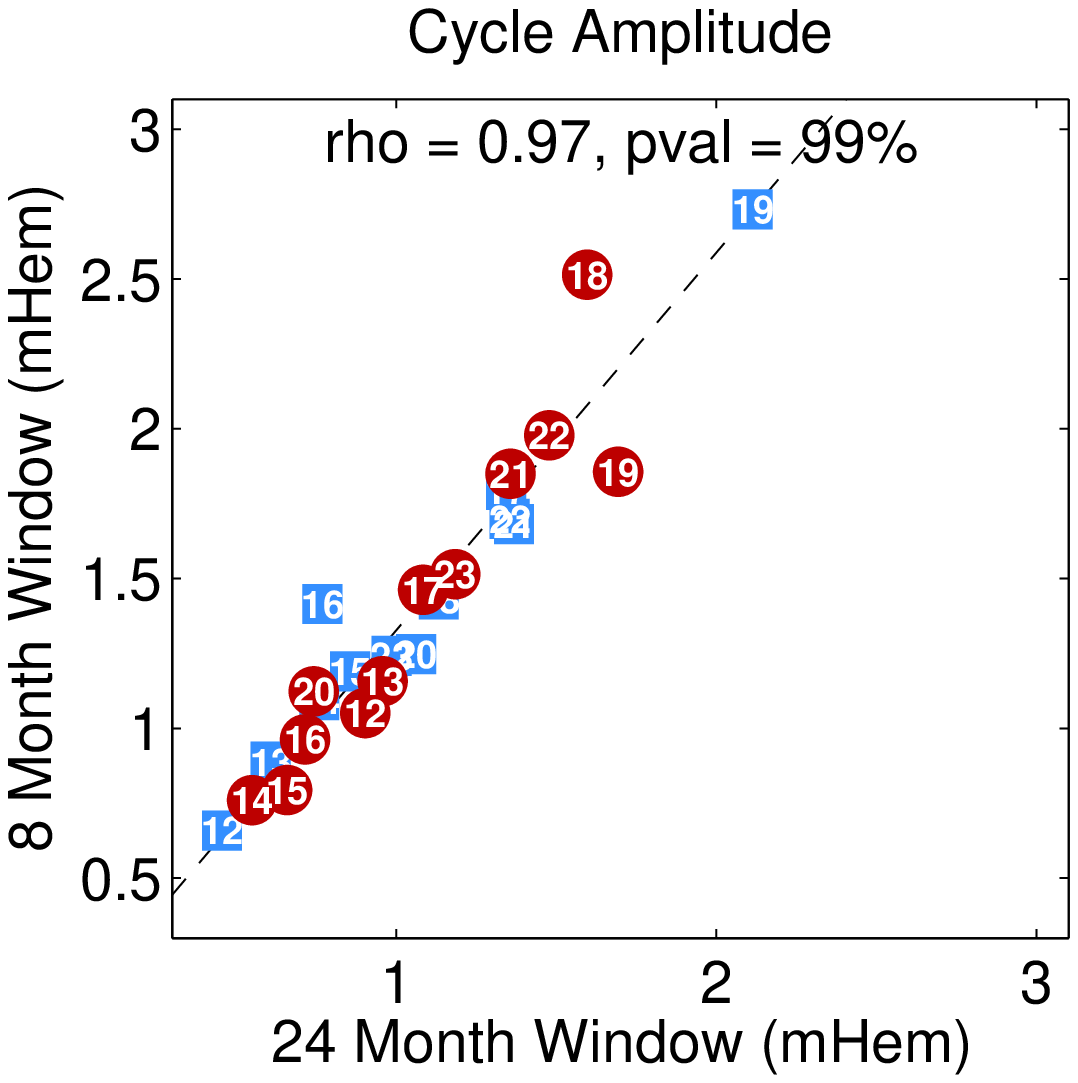}
\end{tabular}
\caption{\emph{Left column:} Scatter plot of width at maximum using a 24-month Gaussian filter vs. 20, 16, 12, and 8 month Gaussian filters. Blue squares and green triangles (red circles and magenta stars) correspond to points in the Northern (Southern) hemisphere.  Points in the main (secondary branch) are denoted with blue and red (green and magenta) markers.  \emph{Middle column:} Smoothed total sunspot area for hemispheric cycle 18 north using 20, 16, 12, and 8 month Gaussian filters.  \emph{Right column:} Scatter plot of cycle amplitude using a 24-month Gaussian filter vs. 20, 16, 12, and 8 month Gaussian filters.}\label{Fig_Filter}
\end{figure}

\section{Dependence of Cycle Properties on Filtering Details}

The conclusions of this work rely on the assumption that maximum amplitude and WaM are independent of changes in cycle filtering; in this section we study how these quantities respond to changes in filter window.  The left column of Figure \ref{Fig_Filter} shows the relationship between WaM calculated using a 24 months window and WaM calculated using 20, 16, 12, and 8 months windows.  We find a strong correlation in all cases (with Spearman rank correlation coefficients between $\rho=0.79$ and $\rho=0.87$ with $99\%$ statistical significance), that is also in good agreement with a $y=x$ relationship.  Of particular importance is the fact that, regardless of the window, the widest maxima are consistently those associated with the secondary branch in the relationship between polar flux at minimum vs.\ the amplitude of the next cycle.  An example of the robustness of this quantity can be found in the middle column of Figure \ref{Fig_Filter} showing how the structure of cycle 18 North changes for different filtering windows (the appearance of cycle 18 North using a 24 month window can be found in Figure~\ref{Fig_Jggnss1}); while using shorter windows introduces a large amount of variability to the shape of the cycle, WaM remains largely unchanged.

The other important quantity related to cycle shape is maximum amplitude, which we also find to be very robust to changes in the size of our filtering window (see right column of Figure \ref{Fig_Filter}).  We find a very strong correlation in all cases (with Spearman rank correlation coefficients between $\rho=0.97$ and $\rho=0.99$ with $99\%$ statistical significance).   Taken together these results suggest that the results presented here are not specific to a choice of filtering details.

\section{Calculation of the Dipolar and Quadrupolar Moments Based on the Northern and Southern Polar Fluxes}

In order to estimate the contribution of the polar flux to the axial dipolar and quadrupolar moments we assume that they are the main determinants of the polar magnetic fields.  Using a potential source surface extrapolation \cite{altschuler-Newkirk1969}, and assuming radial field and axial symmetry, allows us to calculate the magnetic flux inside them as:
\begin{equation}\label{Eq_MM}
   \Phi(\theta_1,\theta_2) = \int_{0}^{2\pi}\int_{\theta_1}^{\theta_2}\left[ 2\operatorname{DM}*\operatorname{P_1}(cos(\theta)) + 3\operatorname{QM}*\operatorname{P_2}(\cos(\theta)) \right]R_{\odot}^2\sin(\theta)d\theta d\phi,
\end{equation}
where $\theta_1=0$ \& $\theta_2=\pi/6$ ($\theta_1=5\pi/6$ \& $\theta_2=\pi$) are the colatitude boundaries of the north (south) pole, $\operatorname{DM}$ and $\operatorname{QM}$ are the dipolar and quadrupolar moments, and $P_n(\cos(\theta))$ are the unassociated Legendre polynomials.  Evaluating the integrals yields
\begin{equation}\label{Eq_MM_PFN}
    \Phi_N = \frac{\pi R_\odot^2}{8}(4\operatorname{DM}+3\sqrt{3}\operatorname{QM})
\end{equation}
for the northern polar flux, and
\begin{equation}\label{Eq_MM_PFS}
    \Phi_S = \frac{\pi R_\odot^2}{8}(-4\operatorname{DM}+3\sqrt{3}\operatorname{QM})
\end{equation}
for the southern polar flux.  Combining these two equations yields
\begin{equation}\label{Eq_DM1}
    \operatorname{DM} = \frac{1}{\pi R_\odot^2}(\Phi_N - \Phi_S)
\end{equation}
and
\begin{equation}\label{Eq_QM1}
    \operatorname{QM} = \frac{4}{3\sqrt{3}\pi R_\odot^2}(\Phi_N + \Phi_S).
\end{equation}

Taking advantage of the fact that polar flux has opposite sign in the northern and southern hemisphere, we refine the dipolar and quadrupolar moments using the absolute value of each polar flux:
\begin{equation}\label{Eq_DM2}
    \operatorname{DM} = \frac{1}{\pi R_\odot^2}( |\Phi_N| + |\Phi_S|)
\end{equation}
and
\begin{equation}\label{Eq_QM2}
    \operatorname{QM} = \frac{4}{3\sqrt{3}\pi R_\odot^2}( |\Phi_N| - |\Phi_S|).
\end{equation}
These expressions have the advantage that the dipolar moment becomes a strictly positive quantity that can be related naturally to sunspot area (also a strictly positive quantity), and that the sign of the quadrupolar moment gives a direct indication of which pole has flux in excess: a positive (negative) quadrupolar moment indicates that the north (south) pole has more flux.  It is also important to note that although this calculation is performed using polar crowns with a latitudinal extent of 30$^o$, using a different value will still result in a dipolar (quadrupolar) moment proportional to the total unsigned polar flux (polar flux imbalance).

\begin{table}[H]
\begin{center}
\begin{tabular*}{\textwidth}{@{\extracolsep{\fill}}c c c c c}
  & \textbf{a}         & \textbf{RMSE} & \textbf{R$^2$} & \textbf{99\% Prediction}\\
  & (mHem/10$^{22}$Mx) & (mHem)        &                & \textbf{Bounds} (mHem)\\
\hline
Main Branch ($a_{mb}$)      & 0.802 & 0.18 & 0.77 & 0.45\\
Secondary Branch ($a_{sb}$) & 0.425 & 0.09 & 0.84 & 0.44\\
\hline
\end{tabular*}
\end{center}
  \caption{Fit parameters of Polar Flux During Minimum vs.\ Amplitude of the Next Cycle (See Equation \ref{Eq_Fit}).  RMSE stands for Root Mean Square Error.}\label{Tab_Fit}
\end{table}

\begin{figure}[b]
\centering
  \includegraphics[width=0.85\textwidth]{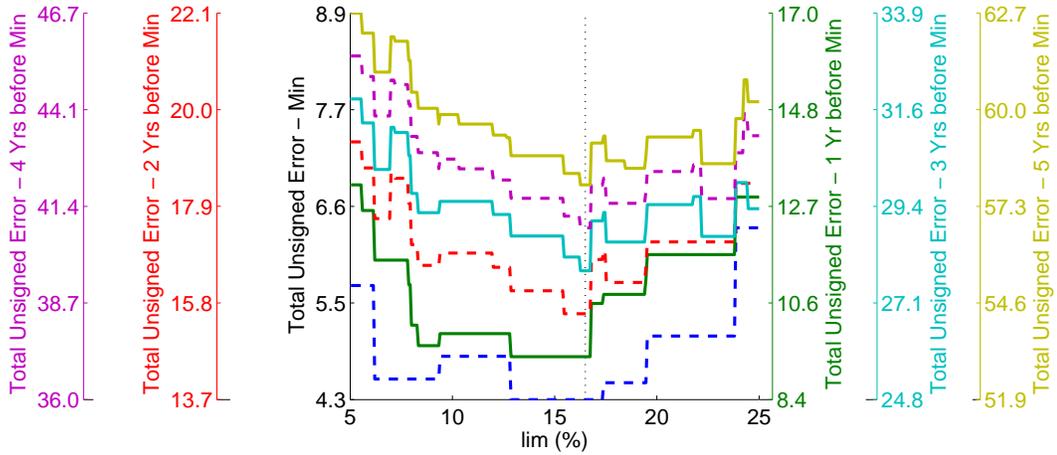}\\
\caption{Total unsigned error (TUE) as a function of the threshold used to determine which of the branches is used for prediction ($lim$).  TUE is calculated using polar flux values at minimum (blue, green, red, cyan, magenta \& yellow), plus the total unsigned error calculated using polar flux values one year (green to yellow), plus two years (red to yellow), plus three years (cyan to yellow), plus four years (magenta \& yellow), plus five years (yellow) before minimum.  Each line has an associated axis, shifted and scaled so that they do not overlap.  A value of $lim=16.5\%$ (vertical black dotted line) minimizes all curves.}\label{Fig_TE}
\end{figure}

\section{Fit Parameters of Polar Flux During Minimum vs.\ Amplitude of the Next Cycle}

We fit each branch using a proportional relationship of the form:
\begin{equation}\label{Eq_Fit}
  y = ax,
\end{equation}
and use the following functions for predicting the amplitude of the next cycle ($n+1$) in the northern hemisphere:
\begin{equation}\label{Eq_Preda}
   \operatorname{Amp}(\operatorname{PFN},lim)_{n+1} =  \left\{\begin{array}{cc} a_{mb}\operatorname{PFN}_n & \frac{\operatorname{QM}}{\operatorname{DM}} \leq lim\\
                                                                            a_{sb}\operatorname{PFN}_n & \frac{\operatorname{QM}}{\operatorname{DM}} > lim
                                                    \end{array}\right.,
\end{equation}
and in the southern hemisphere:
\begin{equation}\label{Eq_Predb}
   \operatorname{Amp}(\operatorname{PFS},lim)_{n+1} =  \left\{\begin{array}{cc} a_{mb}\operatorname{PFS}_n & \frac{\operatorname{QM}}{\operatorname{DM}} \geq -lim\\
                                                                            a_{sb}\operatorname{PFS}_n & \frac{\operatorname{QM}}{\operatorname{DM}} < -lim
                                                    \end{array}\right.,
\end{equation}
where $\operatorname{PFN}_n$ ($\operatorname{PFS}_n$) is the polar flux at the northern (southern) hemisphere at the minimum of cycle $n$, $a_{mb}$ ($a_{sb}$) is the proportionality coefficient of the main (secondary) branch (see Table \ref{Tab_Fit}), $\operatorname{DM}$ ($\operatorname{QM}$) is the dipolar (quadrupolar) moment defined in Equation \ref{Eq_DM2} (\ref{Eq_QM2}), and $lim$ is the threshold on $\operatorname{QM}/\operatorname{DM}$ used to determine which of the branches is used for prediction.

In order to determine $lim$, we minimize the total unsigned error:
\begin{equation}\label{Eq_optim}
    \operatorname{TE}(lim) = \sum_{i=0}^{N_y} \sum_{n=N_0}^{N_c} |\max(A_H)_{n+1} - \operatorname{Amp}(\operatorname{PFH}(i)_n,lim)|,
\end{equation}
where $H$ is used to denote inclusively both the northern and southern hemispheres, the index $n$ indicates each cycle in our dataset, and the index $i$ indicates how many years before minimum are included in the calculation (0 means only estimates using polar flux at minimum, $N_y = 5$ means including estimates using polar flux at minimum, as well as 1-5 years before minimum).  Figure \ref{Fig_TE} shows this function calculated including points from 0 up to 5 years before minimum.  Each line is characterized by a flat global minimum whose extent depends on how many points are included, but which overlaps with the minimum of all other curves.  We find a value of $lim=16.5\%$ to optimize the performance of the algorithm.

\begin{figure}[H]
\begin{center}
\begin{tabular}{ccc}
  \includegraphics[scale=0.37]{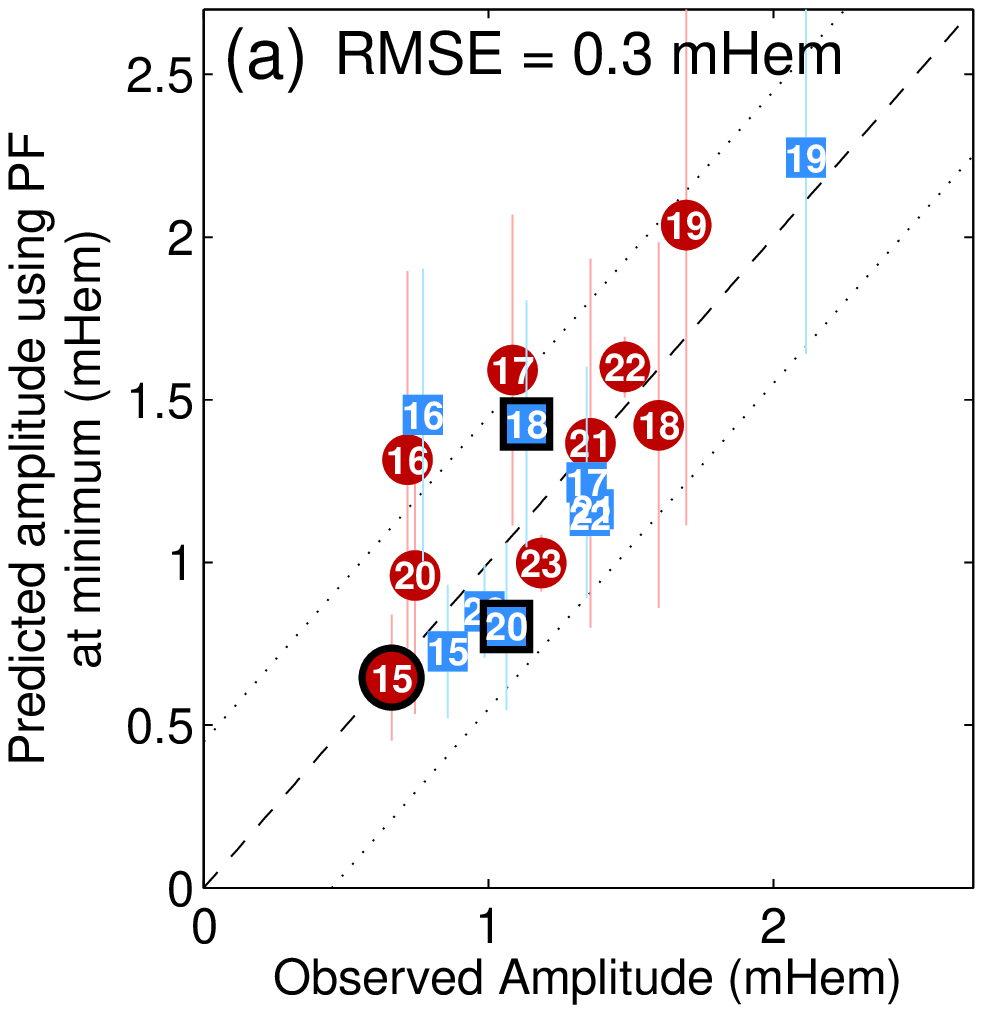} & \includegraphics[scale=0.37]{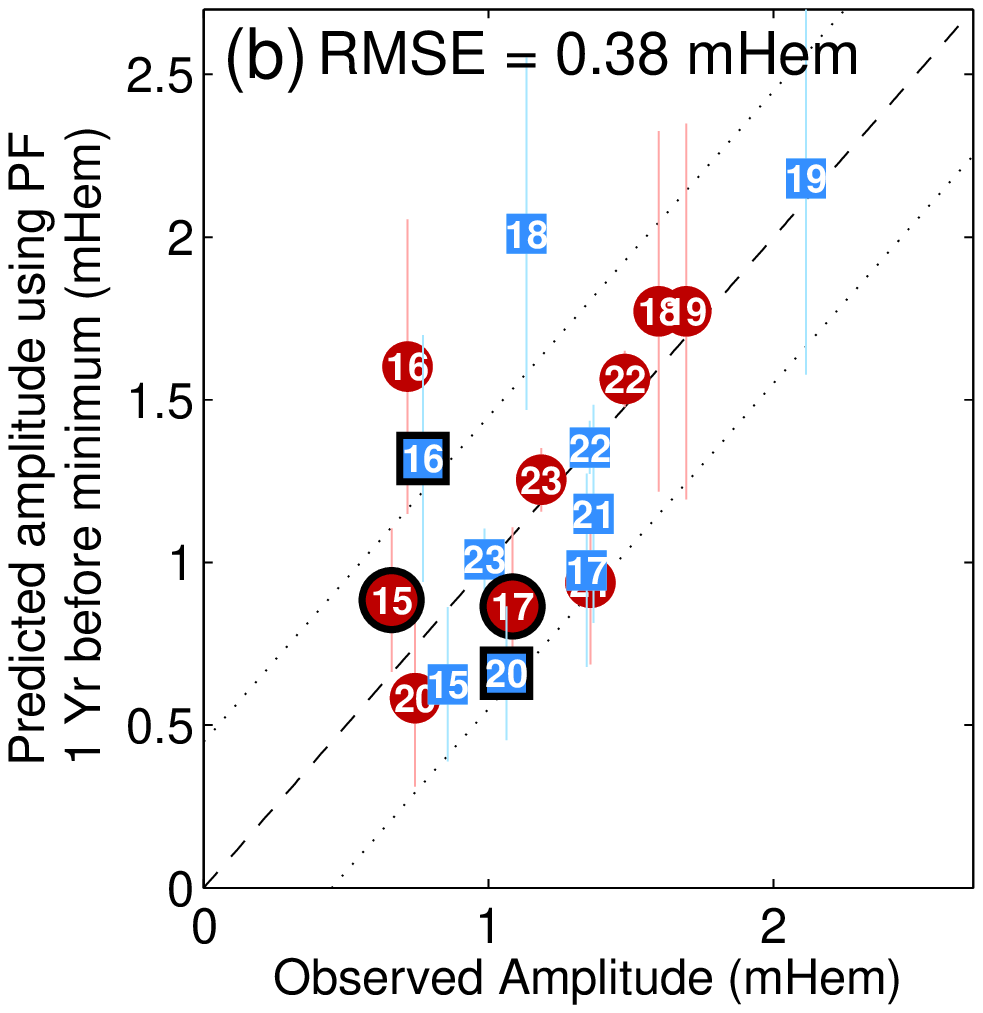} & \includegraphics[scale=0.37]{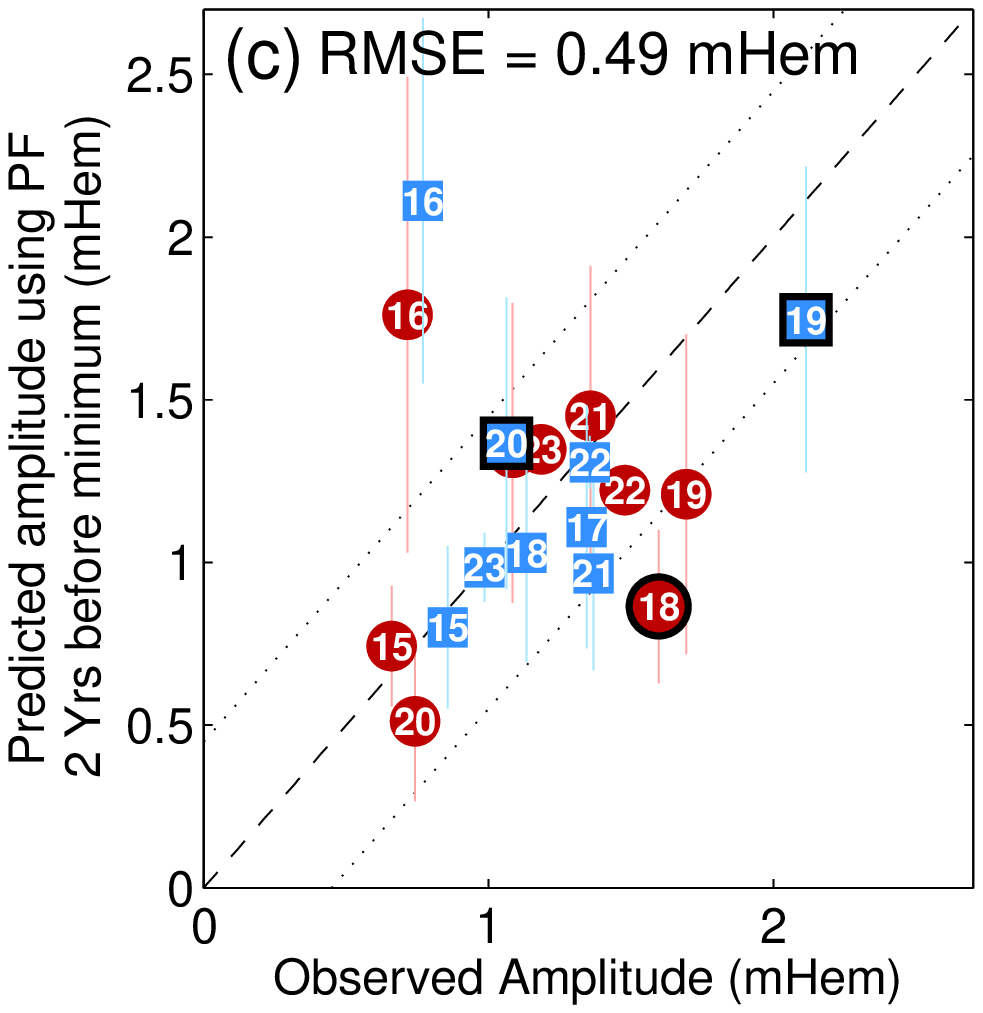} \\
  \includegraphics[scale=0.37]{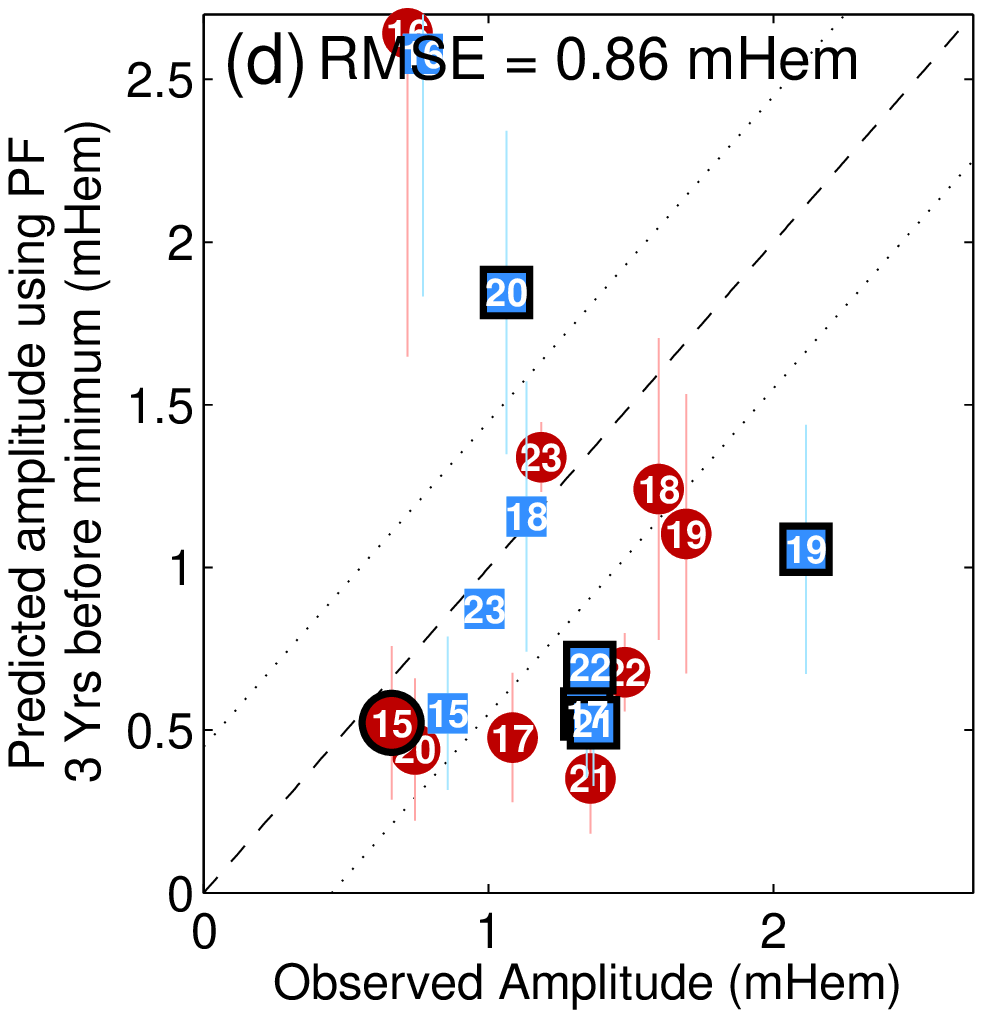} & \includegraphics[scale=0.37]{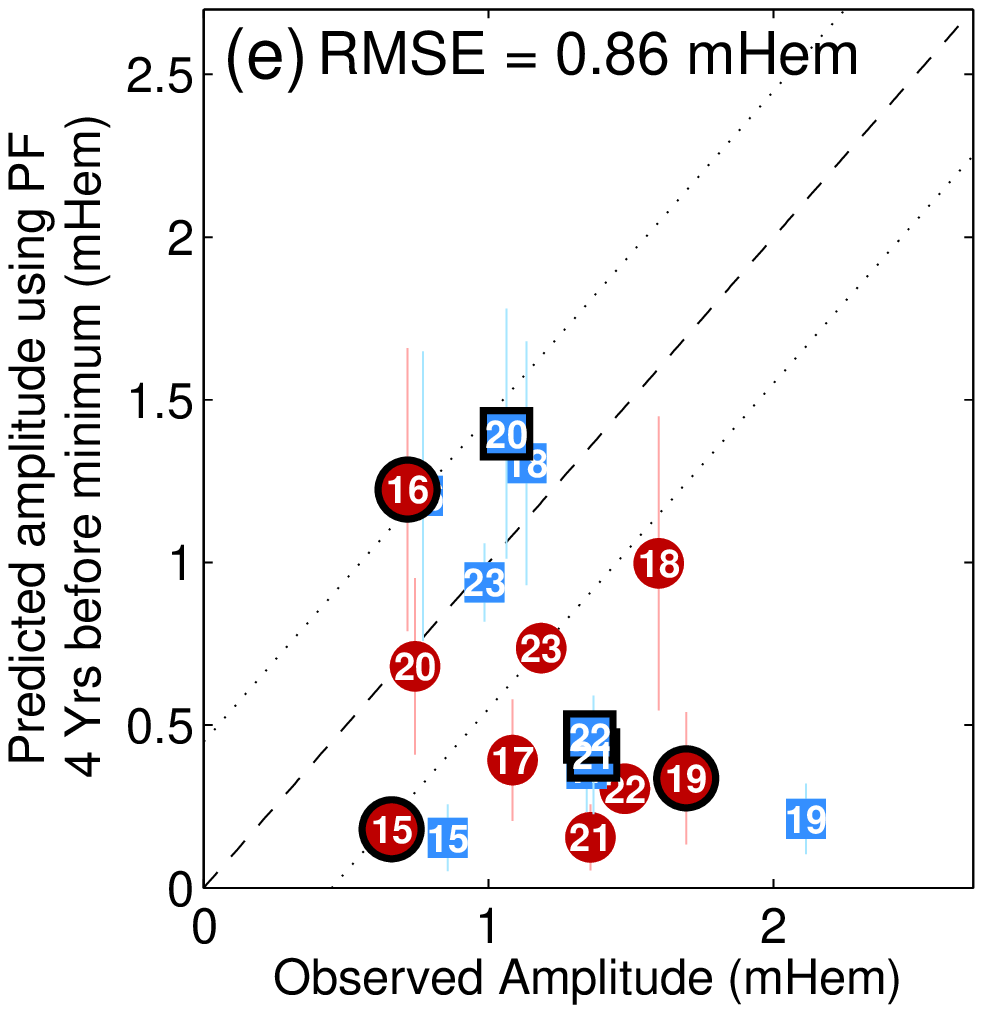} & \includegraphics[scale=0.37]{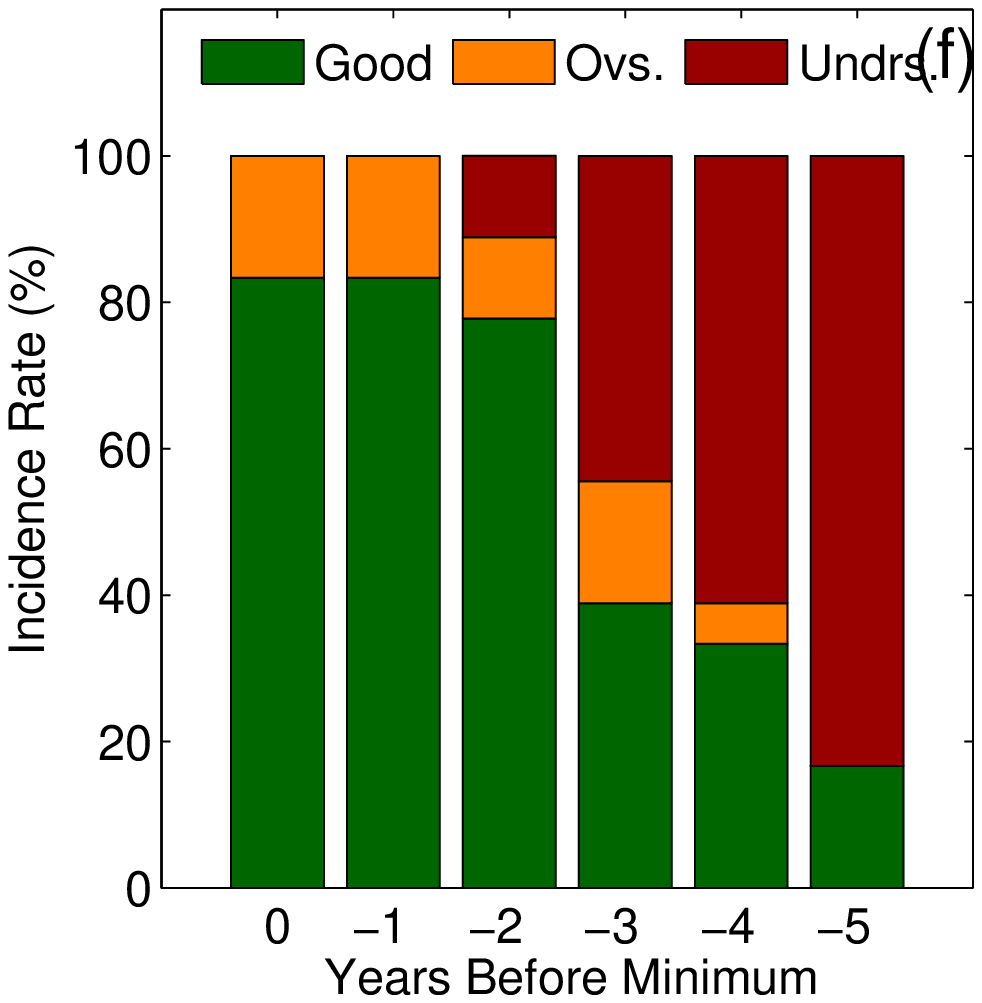}
\end{tabular}
\end{center}
\caption{Predictions using solar minimum conditions (a) and polar flux measurements taken one (b), two (c), three (d), and four (e) years before minimum.  The line $y=x$ is plotted as a dashed line and the $99\%$ confidence bounds as dashed lines.  Blue squares (red circles) correspond to points in the Northern (Southern) hemisphere.  Points with black borders are obtained using the secondary branch relationship.  RMSE corresponds to the root-mean-square error, which is a measure of the average error between the predicted and observed cycle amplitudes. (f) Success rate of the prediction method.}\label{Fig_TimeD}
\end{figure}

\section{Performance of the Cycle Prediction in Time}

The performance of our predictive model can be assessed using plots of polar flux versus amplitude of the next cycle using conditions taken several years before minimum.  Figures \ref{Fig_TimeD}-a to e show model performance plots for all years going from minimum to five years before minimum.  This information is consolidated in the relative performance of the predictive model shown in Figure \ref{Fig_TimeD}-f.  Actual prediction values in mHem and how they compare with observations are tabulated in Table \ref{Tab_Pred}.  It is uncertain why the model does a poor job of predicting cycle 16, both N and S.  However, considering that both hemispheres are among the widest and most irregular at maximum (see Figure \ref{Fig_Jggnss1}), it is likely that higher order magnetic moments are playing an important role in determining the amplitude of the next cycle and thus our method fails.

\begin{table}[H]
\begin{center}
\renewcommand{\arraystretch}{0.75}
\begin{tabular*}{\textwidth}{@{\extracolsep{\fill}}  |c |c |c c c c c c| }
\multicolumn{8}{c}{\textbf{\emph{\large NORTHERN HEMISPHERE}}} \\
\hline
\textbf{CYCLE} & \textbf{OBSERVED} & \multicolumn{6}{c|}{\textbf{PREDICTED}} \\
 &   & \textbf{Min} & \textbf{Min-1Yr.} & \textbf{Min-2Yrs.} & \textbf{Min-3Yrs.} & \textbf{\textbf{Min-4Yrs.}} & \textbf{\textbf{Min-5Yrs.}}\\
\hline
15 & 0.86 & 0.73 & 0.63 & 0.80 & 0.55 & 0.15 & 0.16\\
16 & 0.77 & 1.45 & 1.32 & 2.11 & 2.58 & 1.20 & 0.14\\
17 & 1.34 & 1.25 & 0.98 & 1.11 & 0.55 & 0.37 & 0.06\\
18 & 1.13 & 1.43 & 2.01 & 1.03 & 1.16 & 1.31 & 0.20\\
19 & 2.11 & 2.24 & 2.18 & 1.75 & 1.06 & 0.21 & 0.06\\
20 & 1.06 & 0.80 & 0.66 & 1.37 & 1.85 & 1.40 & 0.15\\
21 & 1.37 & 1.16 & 1.15 & 0.97 & 0.52 & 0.41 & 0.16\\
22 & 1.35 & 1.14 & 1.35 & 1.31 & 0.69 & 0.46 & 0.24\\
23 & 0.99 & 0.85 & 1.01 & 0.99 & 0.87 & 0.94 & 0.33\\
\hline
24 & --  & 0.59 & 0.72 & 0.77 & 0.64 & 0.77 & 0.83\\
\hline
\hline
\textbf{Pred.~Err.}  & -- & 0.24 & 0.30 & 0.32 & 0.71 & 0.71 & 1.05\\
\hline
\multicolumn{8}{c}{ } \\
\multicolumn{8}{c}{\textbf{\emph{\large SOUTHERN HEMISPHERE}}} \\
\hline
\textbf{CYCLE} & \textbf{OBSERVED} & \multicolumn{6}{c|}{\textbf{PREDICTED}} \\
 &   & \textbf{Min} & \textbf{Min-1Yr.} & \textbf{Min-2Yrs.} & \textbf{Min-3Yrs.} & \textbf{\textbf{Min-4Yrs.}} & \textbf{\textbf{Min-5Yrs.}}\\
\hline
15 & 0.66 & 0.65 & 0.88 & 0.74 & 0.52 & 0.18 & 0.23\\
16 & 0.71 & 1.32 & 1.60 & 1.76 & 2.64 & 1.22 & 0.63\\
17 & 1.08 & 1.59 & 0.86 & 1.34 & 0.48 & 0.39 & 0.22\\
18 & 1.60 & 1.42 & 1.77 & 0.86 & 1.24 & 1.00 & 0.41\\
19 & 1.69 & 2.04 & 1.77 & 1.21 & 1.10 & 0.34 & 0.11\\
20 & 0.74 & 0.96 & 0.58 & 0.51 & 0.44 & 0.68 & 0.57\\
21 & 1.36 & 1.37 & 0.94 & 1.45 & 0.35 & 0.15 & 0.11\\
22 & 1.48 & 1.60 & 1.56 & 1.22 & 0.68 & 0.30 & 0.28\\
23 & 1.18 & 1.00 & 1.25 & 1.35 & 1.34 & 0.74 & 0.32\\
\hline
24 & --   & 0.66 & 0.64 & 0.77 & 0.76 & 0.74 & 0.72\\
\hline
\hline
\textbf{Pred.~Err.}  & -- & 0.24 & 0.26 & 0.37 & 0.65 & 0.72 & 0.85\\
\hline\end{tabular*}
\end{center}
  \caption{\textbf{Predicted vs.\ Observed cycle amplitudes.}   Tabulated values of the hemispheric predictions obtained using polar flux measurements at to 1-5 years before solar minimum (shown in the scatter-plots of Figure~\ref{Fig_TimeD}).  All units are given in mHem.  The average prediction error is tabulated in the last row of each hemisphere's table.}\label{Tab_Pred}
\end{table}



\end{document}